\documentclass[aps,pre,twocolumn,floatfix,showpacs,amsmath,10pt]{revtex4-1}
\usepackage{amssymb,graphics}
\usepackage{setspace} 
\usepackage{graphicx}
\usepackage{hyperref}
\usepackage{xcolor}
\newcommand{\w}{w}

\newcommand{\showfigures}[1]{{#1}} 

\begin{document}

\title{Growth, collapse, and stalling in a  mechanical model for neurite motility}
\author{Pierre Recho$^{1}$}
\email{pierre.recho@polytechnique.edu}
\author{Antoine Jerusalem$^2$} 
\email{antoine.jerusalem@eng.ox.ac.uk}
\author{Alain Goriely$^1$}
\email{goriely@maths.ox.ac.uk}

\affiliation{$^1$ Mathematical Institute, University of Oxford,
Oxford OX26GG, United Kingdom}
\affiliation{$^2$ Department of Engineering Science, University of Oxford,
Oxford OX13PJ, United Kingdom}

\date{\today}

\begin{abstract}
Neurites, the long cellular protrusions that form the routes of the neuronal network are capable to actively extend during early morphogenesis or to regenerate after trauma. To perform this task, they rely on their cytoskeleton for mechanical support. In this paper, we present a three-component active gel model that describes neurites in the three robust mechanical states observed experimentally: collapsed, static, and motile. These states arise from an interplay between the physical forces driven by growth of the microtubule-rich inner core of the neurite and the acto-myosin contractility of its surrounding cortical membrane. In particular, static states appear as a mechanical traction/compression balance of these two parallel structures. The model  predicts how the response of a neurite to a towing force  depends on the force magnitude and recovers the response of neurites to several drug treatments that modulate the cytoskeleton active and passive properties. 
\end{abstract}
\maketitle

\section{Introduction}

Neurons are cells with long and thin ($\sim 1\mu $m in diameter) quasi one-dimensional processes called neurites, a term which comprises the axon which emits electric signals and dendrites which are generally shorter and receive signals. These processes emerge from the cell body (the \emph{soma}) and, during embryonic development or regeneration after a trauma, are able to crawl over large distances to reach targets from other neurons thus forming a complex nervous network essential for both perception and motion. Understanding how neurites can establish these long distance connections is a problem which was pioneered more then a century ago \cite{Cajal1911} and is of paramount therapeutic importance. For instance, injuries of the spinal cord are often characterised by an irreversible and debilitating loss of motor and sensory functions of the lower body (paraplegia, tetraplegia) because disrupted neurites cannot overcome the inflammation and are incapable to initiate extensions that would rebuilt the broken connections \cite{Silver2004, Case2005, Deumens2010}.

The cytoskeleton of neurites is the mechanical scaffold that maintains their morphology and motility \cite{Mitchison1988, Dent2003, Franze2010, Coles2015}. Extrinsic and intrinsic guidance clues may be viewed as agents that influence the physical state of the cytoskeleton via biochemical pathways  \cite{Aeschlimann2001,Dickson2002, Lowery2009, Reingruber2014}. For example, the concentration of calcium is known to influence the Rho pathway which in turn modulates the neurite contractility and can lead to a reversible collapse that shortens the axon \cite{Franze2009}. 

\begin{figure}[!t] 
\begin{center}
\showfigures{
\includegraphics[scale=0.5]{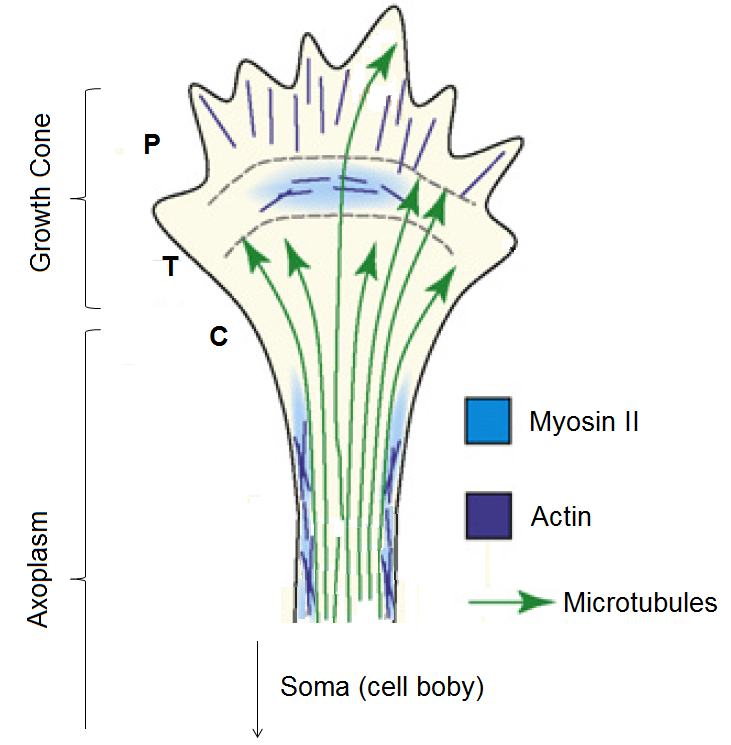}
}
\caption{
\label{fig:axon_scheme} Schematic representation of the cytoskeleton of a neurite extending from the soma. Adapted from Ref. \cite{Vallee2009}. Colors online.}
\end{center}
\end{figure}

The neuronal cytoskeleton (see Ref. \cite{Coles2015} for an extensive review and Fig.~\ref{fig:axon_scheme} for a simplified scheme) is a meshwork of three main types of biological polymers: F-actin, microtubules, and neurofilaments which all contribute mechanically \cite{Ouyang2013}. While neurofilaments are passive and apolar, both F-actin and microtubules are capable to polymerise at one end (addition of G-actin and tubulin subunits, respectively) and depolymerise at the other (by removal of subunits) with potentially low ($\sim 1$ min) turnover duration. They can also both be cross-linked by molecular motors (myosin II for F-actin, dynein or kinesin for microtubules) that are able to exert active stresses inside the meshwork \cite{Brown2003, Roossien2014, Lu2013}. Following Ref. \cite{Mitchison1988}, we define two different compartments of the neurite where the cytoskeleton is organised in a different way: the \textit{kinetoplasm}, or growth cone (GC), at the proximal end of the neurite and the \textit{axoplasm} which connects the GC to the soma. The axoplasm contains a core array of para-axial microtubules connected by passive cross-linkers (MAP: Microtubules Associated Proteins) that generate a network with a quasi-lattice structure \cite{Ahmadzadeh2014}. These microtubules are highly stable (turnover duration of hours) possibly due to the presence of MAPs. In the axoplasm, F-actin is mostly organised into a cortical mesh around the microtubule's inner core and the presence of myosin II motors in this cortex leads to the presence of contractile stresses \cite{Julicher2007}. These two meshworks are physically connected by different types of special proteins (such as +TIP, see Ref. \cite{Coles2015}) which mediate force transmission between them. In continuity with this cortex, the GC is almost free of microtubules, apart from those engaging into filopodia while F-actin is organised in a lamellipodium similar to the ones found in cells specialised in crawling (such as keratocytes \cite{Verkhovsky1999}). Filaments polymerising at the leading edge (the \textit{P-domain}) protrude the membrane and are then advected backward by a retrograde flow powered by myosin II motors that concentrate at the trailing edge of the GC (the \textit{T-domain}) \cite{Medeiros2006}. Actin then accumulates into thick bundles in the T-domain that constitutes the main obstacle preventing microtubules from entering the GC. The cytoskeleton is connected to the external substrate/cellular matrix by special proteins specialised in adhesion such as integrins and cadherins \cite{Bard2008, Chan2008}. 
 
Numerous authors have proposed physical models to explain how the neurite cytoskeleton drives its motility \cite{Aeschlimann2000, Kiddie2005, Suter2011, Franze2010}. These models can be divided into two main classes depending on whether they imply that the GC pulls the trailing axoplasm thanks to F-actin polymerisation pushing the membrane in the P-domain \cite{Peskin1993, Mogilner2003} and myosin II contractility pulling from the T-domain \cite{Bray1979, Lamoureux1989, OToole2008, OToole2015}, or whether it is rather microtubules which, from the axoplasm, polymerise against the T-domain and propel the GC \cite{Van1994, Rauch2013}, or both \cite{Garcia2012}. As the rate of polymerisation of microtubules depends on the force applied at their tip \cite{Dogterom1997}, both effects have been considered in a single simple model \cite{Buxbaum1992}. Experimentally, the physical forces arising from the F-actin or the microtubule meshworks are important  since drug treatments lowering myosin II contractility (blebbistatin), preventing actin polymerisation (cytochalasin B), depolymerisating microtubules (nocodazole) or on the contrary stabilising them (taxol) each influence the tip velocity of a crawling neurite in a concentration dependent manner \cite{Letourneau1987, Dennerll1988, Brown2003, Kollins2009, Vallee2009, Hur2011}. Interestingly, different levels of cytochalasin B can lead to either an increase \cite{Dennerll1988} (high level) or a decrease \cite{Zheng1996} (low level) of the speed. This dependence suggests an antagonistic role of the acto-myosin meshwork in the propulsion.  

Mechanical models require a rheological characterisation of the axoplasm and the GC. The axoplasm has been described as a Burgers viscoelastic material based on its fast elastic response (seconds to minutes) to a force applied laterally or at the tip   \cite{Dennerll1989, Bernal2007, Bernal2010} while it elongates at a constant rate on longer time scales (a few hours) in response to a constant high force \cite{OToole2008, OToole2015}. In Ref. \cite{OToole2008, OToole2015}, experiments tracking the mitochondria docked on the microtubule array have shown that, close to the T-domain, this network flows forward with a velocity comparable to the velocity of the neurite. Away from the T-domain towards the soma, the velocity  decays exponentially, suggesting a fluid-like behaviour of the neurite. However, if the applied force is not large enough, the neurite may undergo a finite deformation instead of acquiring a finite velocity \cite{Zheng1991} and both rheological models of Ref. \cite{Dennerll1989} and \cite{Bernal2010} indicate a long term stiffness of the neurite two orders of magnitude smaller than the short time one. Furthermore, the loading rate is also known to play an important role in the possible action potential impairment of the axon or in the transport properties alteration resulting from a loss of connectivity of the microtubules network \cite{Jerusalem2014, Ahmed2014, Ahmadzadeh2014, Shamloo2015}. Dynamic loading is not studied in this article and the loading is assumed to be quasi-static. Axoplasm active growth and contractility have been proposed as the potential drivers for retraction or elongation of the neurite in presence of an applied force  \cite{Dennerll1989}, and  contractility is explicitly incorporated as a force opposing elongation in Ref. \cite{Bernal2010, OToole2015}. 

The GC has been characterised as a Maxwell viscoelastic fluid with a relaxation time of a few seconds and an active contractile pre-stress stemming from the motor activity at the rear of the cone \cite{Betz2011}. The F-actin polymerisation driven formation of filopodia extension and retraction has been physically described in Ref. \cite{Betz2006, Craig2012}. 

In the present paper, we follow the suggestion of Ref. \cite{OToole2015} that the theory of active gel may be used to unify these aforementioned models in order to obtain a global picture of neurite motility. Our one-dimensional model is based on the particular geometry of the neurite cytoskeleton and fundamental balance laws. An analysis of its solutions reveals that, depending on the neurite passive and active rheology, the neurite can collapse to the soma, remain static, or grow at finite velocity. The interchangeability of these three states is consistent with experiments that modify the state of the cytoskeleton and its substrate adhesivity with drugs.  

In particular, we recover within a common framework the following general trends emerging from different sets of experiments probing the mechanical and structural environment of growing neurites: \begin{itemize}
\item \textbf{Growth under axial force:} As mentioned above and already observed 30 years ago \cite{Bray1979}, a steady axial force applied by a cantilever at the proximal tip of a neurite elongates it \cite{Zheng1991, Dennerll1989, Lamoureux1989, Lamoureux2002, OToole2008, Nguyen2013, OToole2015}. This elongation is elastic if the force is below a certain threshold. Above that threshold, the neurite grows with finite velocity \cite{Heidemann1993}. 
 \item \textbf{Retraction under microtubules depletion:} As shown in Ref. \cite{Ahmad2000}, neurites retract in response to microtubles depletion and elongate even \emph{in vivo} following stimulation of microtubules polymerisation \cite{ruschel2015systemic}.
 \item \textbf{Retraction with reduction of adhesivity:} By culturing neurites on different substrates, it was shown that retraction is promoted when the substrate adhesivity is reduced \cite{OToole2015}.

\item \textbf{Motility is related to contractility:}  Finally, neurites initiate their motility in a robust way when exposed to drugs that impair their active contractility \cite{Hur2011, Yu2012}. 
\end{itemize}
The last two effects \cite{Ketschek2007} are particularly important as possible therapeutic targets to promote axon regeneration after trauma \cite{Yu2012}.  

The paper is organised as follows: In Section \ref{Axoplasm_ propulsion}, we develop a mechanical model for the axoplasm (acto-myosin and microtubules phases) alone and study its motility properties under an applied proximal traction force. In Section \ref{gc_propulsion}, we model the GC (acto-myosin phase only) motile properties in response to a traction force applied at the trailing edge. In Section \ref{full_neurite}, we combine both models by assuming stress continuity at the T-domain to obtain a complete model of a growing neurite and show that it compares well with experiments.

\section{Axoplasm propulsion}\label{Axoplasm_ propulsion}

\begin{figure}[!t] 
\begin{center}
\showfigures{
\includegraphics[scale=0.45]{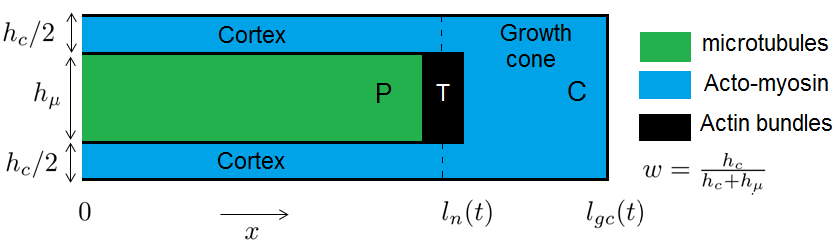}
}
\caption{
\label{fig:mecha_scheme} Schematic of the neurite model geometry. Colors online}
\end{center}
\end{figure}

Following Ref. \cite{Holland2014}, we model the microtubule network core of the axoplasm as a one-dimensional morphoelastic rod whose material points are indexed by the coordinate $x\in[0,l_n(t)]$, $0$ denoting the connection with the soma and $l_n(t)$ the moving boundary between the GC and the axoplasm (T-domain) as shown in Fig.~\ref{fig:mecha_scheme}. Notice our model does not separate the contribution of the neurofilamants from the one of the microtubules. They are thus viewed as a passive reinforcing structure contributing to the overall network elasticity \cite{Ouyang2013}.
This rod can only deform along the axis and is in frictional contact  with a viscous contractile ``sleeve'' (the cortex) which is supported by a static background. The tip of the neurite is subjected to a given traction force exerted either by the GC in normal growth conditions or  by a micropipette, in experiments where the GC is lifted from the substrate~\cite{OToole2015}.

\subsection{Balance of mass}
\paragraph*{Microtubule network}
Let $\rho_{\mu}$ denote the mass density of microtubules and $v_{\mu}$ their velocity in the lab reference frame. The mass balance equation reads 
\begin{equation}\label{massbal}
\partial_t\rho_{\mu}+\partial_x(\rho_{\mu}v_{\mu})=S_{\mu},
\end{equation}
where the  source term  
\begin{equation}\label{sourceterm}
S_{\mu}=k_p^{\mu}-k_d^{\mu}\rho_{\mu},
\end{equation}
follows a first-order kinetic with a {polymerisation} rate $k_p^{\mu}$ and a {depolymerisation term} $k_d^{\mu}\rho_{\mu}$, proportional to the density \cite{prost2015active}.
We assume here that the tubulin (microtubule subunits) concentration is homogeneous \cite{Bressloff2015} along the neurite because its motor driven transport is much faster ($\sim 1\mu\text{m}.\text{s}^{-1}$) than the crawling velocity ($\sim 10\mu\text{m}.\text{h}^{-1}$). We can rewrite Eq.\eqref{sourceterm} as $S_{\mu}=(\bar{\rho}-\rho_{\mu})/\tau$ where $\bar{\rho}=k_p^{\mu}/k_d^{\mu}$ is the density at chemical equilibrium and $\tau=1/k_d^{\mu}$ is the turnover timescale associated with microtubule renewal.  Here we have adopted a mean field description of the network and we do not consider the microtubule polarity. Indeed, while this information is likely to be important for transport properties along the neurite, microtubules have a clear forward polarity in the axon and a mixed one in dendrites \cite{Coles2015}. Yet, these two structures can equivalently move, suggesting that polarity may not be a fundamental variable in this physical process. Also note that while we do not account for the influence of a potential loading on the kinetic rates  $k_p^{\mu}$ and $k_d^{\mu}$, our mean field description captures the load-dependant dynamic of the whole microtubule array (see Section \ref{solflu}).

Assuming that no microtubule  comes from the growth cone,
Eq.~\eqref{massbal} is equipped with a no-flux boundary condition at the tip of the neurite 
\begin{equation}\label{noflux}
\dot{l}_n=v_{\mu}(l_n(t),t).
\end{equation}
For simplicity, we also impose a no-flux boundary condition at the connection with the soma, so that  $v_{\mu}(0,t)=0$.  
Notice, that at both ends, there is no assumption on the flux of tubulin which adjusts to maintain a constant concentration as hypothesised in \eqref{sourceterm}.

\paragraph*{Cortical network}
Denoting $\rho_c$ the mass density of actin in the cortex, we can write an conservation equation similar to Eq. \eqref{massbal}: 
\begin{equation}
\partial_t\rho_{c}+\partial_x(\rho_{c}v_{c})=S_{c},
\end{equation}
where $v_c$ is the velocity of the actin network in the cortex in the lab reference frame. However, we assume that the actin network is highly compressible compared to the microtubule network. Therefore this equation decouples from the rest of the system and actin density can be found post-factum when the velocity field $v_c$ is known using the method of characteristics \cite{Recho2013b}. This point is not tackled in the present paper and $S_{c}$ is thus left unspecified. 

We again assume that there is no filamentous actin flux from the soma to the cortex: $v_c(0,t)=0$. Unlike the microtubule network, the cortical actin is not stopped at the T-domain and can flow freely in the growth cone. Thus, there is no condition such as Eq. \eqref{noflux} for the cortical flow.

\subsection{Balance of linear momentum}

\paragraph*{Microtubule network}
The microtubule network is in contact with the cortical actin network through different types of cross-linking proteins that can actively bind and unbind (see Ref. \cite{Coles2015} for a review).  Assuming a sufficiently fast binding/unbinding dynamic \cite{Tawada1991}, we  model this contact as a viscous friction. Neglecting inertia, the  balance of linear momentum reads
\begin{equation}\label{forcebalmicro}
\partial_x\sigma_{\mu}=\zeta_{\mu}(v_{\mu}-v_c),
\end{equation}
where $\zeta_{\mu}$ is a friction coefficient and $\sigma_{\mu}$ is the internal axial stress rescaled by the microtubule network width, that is, if $h_c$ is the width of the cortex and $h_{\mu}$ the width of the microtubule network both assumed to be constant, then $\sigma_{\mu}=(1- \w)\Sigma_{\mu}$, where $\Sigma_{\mu}$ is the axial Cauchy stress (axial force per unit  area). We denote 
$$ \w={h_{c}\over h_{\mu}+h_{c}}\in\ [0,1]$$ as the ratio of the cortical over total width of the axon. 

At the leading edge, the axoplasm is subjected to a traction stress $Q$. Thus, the boundary condition associated to Eq. \eqref{forcebalmicro} reads
$$\sigma_{\mu}(l_n(t),t)=(1- \w)Q.$$

\paragraph*{Cortical network}
Similarly, the linear momentum balance in the cortical layer reads 
\begin{equation}\label{forcebalcor}
\partial_x\sigma_{c}=-\zeta_{\mu}(v_{\mu}-v_c)+\zeta_c v_c, 
\end{equation}
where $\zeta_c$ is a friction coefficient of the cortex with respect to the substrate through adhesive proteins \cite{Bard2008}.  Here, $\sigma_{c}= \w\Sigma_{c}$ is the rescaled axial stress, so that the boundary condition at the leading edge is,
$$\sigma_{c}(l_n(t),t)= \w Q.$$

\subsection{Constitutive relations}
To close our system of equations we posit two assumptions about the rheology of the microtubules and cortical networks. 

\paragraph*{Microtubule network}
Given the long turnover time of microtubules inside the axoplasm which are highly stabilised \cite{Mitchison1988, Coles2015} and their high stiffness compared to F-actin filaments, we consider this network to be elastic for the timescale of interest (hours). 
We assume a logarithmic elastic stress-strain dependence
\begin{equation}\label{behavmicro}
\sigma_{\mu}=-(1- \w)E\log\left(\frac{\rho_{\mu}}{\rho_0}\right),
\end{equation}
which has the advantage to prevent both infinite contraction and dilution. The natural density of microtubules at which no stress is created is $\rho_0$. In principle, this density can be modulated by the presence of molecular motors in the microtubule array \cite{OToole2015}. It is worth mentioning that our results are robust with respect to the choice of other increasing concave function than $\log$. In fact, realistic parameters show that $\rho_{\mu}$ is rather close to $\rho_0$ thus implying that in the range of strain considered in physiological conditions, a linear relation between the stress and the  local density (representing the strain in 1D)  could potentially be acceptable too.

\paragraph*{Cortical network} The turnover duration of an actin fibre in the cortex is fast (few seconds \cite{Betz2011, Coles2015}) and we use a linear viscous law for this phase to relate the  stress to the strain rate. In addition, we assume that there is  an active contractile stress created by the myosin II motor activity \cite{Julicher2007}: 
\begin{equation}\label{behavcor}
\sigma_{c}= \w (\eta\partial_x v_c+\chi c).
\end{equation}
The bulk viscosity of F-actin is $\eta$, $\chi>0$ is the contractility coefficient and $c$ the concentration of motors. The conservation equation for $c$ is (see Ref. \cite{Kruse2005, Hannezo2015} for further details),
\begin{equation}\label{mot_bal}
\partial_t c+\partial_x( cv_c)-D\partial_{xx}c=\frac{\bar{c}-c}{\tau_c}.
\end{equation}
The motors are advected with the actin filament that they bind  but can also thermally diffuse with a diffusion coefficient $D$. The linear reaction term accounts for the attachment/detachment of motors with a cycle time $\tau_c$. The concentration of motors at chemical equilibrium is $\bar{c}$. We can supplement this equation with a no-flux boundary condition at the soma, $\partial_xc(0,t)=0$ and assume that the motor concentration in the T-domain is a constant $c(l_n(t),t)=c_0$.

Note that we do not resolve the radial component of the stress in our model (similarly to a neurite constrained in a channel \cite{Tomba2014}). Radial stress will be important for further investigation on the shape and turning of neurites which is outside the scope of this paper.

\subsection{Final system}

Denoting $\sigma=\sigma_c+\sigma_{\mu}$ the total stress, $v=v_c$ the velocity of actin and $\rho=\rho_{\mu}$ the density of microtubules, combining our model equations, we obtain the final system
\begin{equation}\label{Model_final}
\left\lbrace \begin{array}{l}
-\frac{ \w \eta}{\zeta_c}\partial_{xx}\sigma+\sigma= \w \chi c-E(1- \w)\log\left( \frac{\rho}{\rho_0}\right)\\
\partial_t\rho+\partial_x(\rho v)-\frac{E(1- \w)}{\zeta_{\mu}}\partial_{xx}\rho=\frac{\bar{ \rho}-\rho}{\tau}\\
\partial_tc+\partial_x(c v)-D\partial_{xx}c=\frac{\bar{c}-c}{\tau_c},
\end{array}\right. 
\end{equation}
where the velocity field is related to the stress by $v=\partial_x\sigma/\zeta_c$. The boundary conditions are:
\begin{equation}\label{Model_final_BC}
\left\lbrace \begin{array}{l}
\partial_x\sigma\vert_{0}=0\text{, }\partial_x \rho\vert_{0}=0\text{ and }\partial_x c\vert_{0}=0\\
\sigma\vert_{l_n}=Q\text{, }\rho\vert_{l_n}=\rho_0\text{e}^{-Q/E} \text{ and }c\vert_{l_n}=c_0.
\end{array}\right. 
\end{equation}
The last no-flux boundary condition,
\begin{equation}\label{Stefan_condition}
\dot{l}_n=v\vert_{l_n}-\frac{E(1- \w)}{\zeta_{\mu}}\frac{\partial_x\rho}{\rho}\vert_{l_n(t)},
\end{equation}
is a Stefan condition needed to compute the unknown time dependence of the free boundary $l_n(t)$.  
In general,  initial conditions should also be given but here,  we focus on steady states only.  Equation $(\ref{Model_final})_2$ is obtained by expressing the velocity $v_{\mu}$ using Eq.\eqref{forcebalmicro} and combining it with Eq.\eqref{massbal}. 

We comment on the structure of Eq. \eqref{Model_final}: the stress is created  non-locally over the so-called hydrodynamic length $l_c=\sqrt{\eta/\zeta_c}$ by two active agents. Motors from the cortex  are pullers creating a contractile stress and microtubules, provided their density is larger than $\rho_0$, are pushers creating a tensile stress due to the addition of tubulin subunits in the network (growth). If tubulin subunits are removed (shrinking), the microtubules are also pullers. In the context of cell motility, such structures with pushers and pullers have been investigated in Ref. \cite{Carlsson2011, Recho2015a} where it was shown that growing and contracting agents can conspire to achieve robust motile properties of an active segment. Here, we supplement the picture with the two simple and similar dynamic Eq. $\text{(\ref{Model_final})}_{2,3}$ ruling the distribution of pushers and pullers which are relevant in the case of axonal motility.

Having already investigated the pullers dominated case in Ref. \cite{Recho2015b} for motility properties and for the formation of periodic  F-actin rings \cite{Hannezo2015} which are actually observed in axons \cite{Xu2013, Coles2015}, we turn our attention to the pushers dominated case. To understand this regime, here and subsequently, we restrict our attention to the case where  the concentration of motors is homogeneous in the cortex, i.e., $c\equiv \bar{c}$. The resulting system can then be rewritten in the following minimal form,
\begin{equation}\label{Model_final_sim}
\left\lbrace \begin{array}{l}
 \w \eta\partial_{xx}v-\zeta_cv=E(1- \w)\frac{\partial_x\rho}{\rho}\\
\partial_t\rho+\partial_x(\rho v)-\frac{E(1- \w)}{\zeta_{\mu}}\partial_{xx}\rho=\frac{\bar{ \rho}-\rho}{\tau}\\
\end{array}\right. 
\end{equation}
 with boundary conditions,
\begin{equation}\label{Model_final_BC_sim}
\left\lbrace \begin{array}{l}
v\vert_{0}=0\text{ and }\partial_x \rho\vert_{0}=0\\
\eta\partial_xv\vert_{l_n}=Q-\chi \bar{c}\text{ and }\rho\vert_{l_n}=\rho_0\text{e}^{-Q/E},
\end{array}\right. 
\end{equation}
along with the Stefan condition in Eq. \eqref{Stefan_condition}. Substituting the non-dimensional quantities,
\begin{equation}\label{non_dim}
\tilde\sigma=\frac{\sigma}{E}\text{, } \tilde x=\frac{x}{\sqrt{\eta/\zeta_{\mu}}}\text{, } \tilde t=\frac{t}{\eta/E}, \tilde\rho=\frac{\rho}{\overline{\rho}}, \text{ and } \tilde Q=Q/E,
\end{equation}
in Eq. (\ref{Model_final_sim}-\ref{Model_final_BC_sim}) and dropping the tildes for clarity, we obtain the system
\begin{equation}\label{Model_final_sim_nd}
\left\lbrace \begin{array}{l}
 \w\partial_{xx}v-av=(1- \w)\frac{\partial_x\rho}{\rho}\\
\partial_t\rho+\partial_x(\rho v)-(1- \w)\partial_{xx}\rho=\epsilon(1-\rho)
\end{array}\right. 
\end{equation}
 with boundary conditions,
\begin{equation}\label{Model_final_BC_sim_nd}
\left\lbrace \begin{array}{l}
v\vert_{0}=0\text{ and }\partial_x \rho\vert_{0}=0\\
\partial_xv\vert_{l_n}=Q-Q_{c}\text{ and }\rho\vert_{l_n}=\text{e}^{Q_{\mu}-Q}
\end{array}\right. 
\end{equation}
and the free boundary condition,
\begin{equation}\label{Stefan_condition_nd}
\dot{l}_n=v\vert_{l_n}-(1- \w)\frac{\partial_x\rho}{\rho}\vert_{l_n(t)}.
\end{equation} 
We have  now six non-dimensional parameters: 
\begin{itemize}
\item $\w$, the relative width of the cortex with respect to the microtubule network;
\item $a=\zeta_c/\zeta_{\mu}$, the ratio of friction coefficients;
\item $\epsilon=\eta/(\tau E)$, the ratio of the acto-myosin over microtubule network viscosities;
\item $Q$, the applied load at $l_n$ scaled by $E$;
\item $Q_{c}=\chi \bar{c}/E>0$, the scaled contractile load;
\item  $Q_{\mu}=-\log(\bar{\rho}/\rho_0)$ (see below).
\end{itemize}
 Note that these six non-dimensional parameters could be reduced to five by defining $\hat{Q}=Q-Q_{\mu}$ and $\Delta Q_{\mu}^c=Q_{c}-Q_{\mu}$. However, to keep the treatment of the microtubule and  acto-myosin meshwork parallel, we  keep the three distinct loads.

The system of Eq. (\ref{Model_final_sim_nd}-\ref{Stefan_condition_nd}) cannot be explicitly solved but some asymptotic cases give insight on how the physics of such medium works.

\subsection{``Solid'' and ``fluid'' asymptotic cases}\label{solflu}
\paragraph*{The no-microtubule case, $ \w\rightarrow 1$}
In the absence of microtubules, the velocity field can be solved directly from Eq. $\text{(\ref{Model_final_sim_nd})}_1$ and we obtain,
$$v(x,t)=\frac{Q-Q_{c}}{\sqrt{a}}\frac{\sinh(\sqrt{a}x)}{\cosh(\sqrt{a}l_n(t))}.$$
Plugging this expression in Eq. \eqref{Stefan_condition_nd} leads to
$$\dot{l}_n(t)=\frac{Q-Q_{c}}{\sqrt{a}}\tanh(\sqrt{a}l_n(t)).$$
This case was investigated in the absence of contraction ($Q_{c}=0$) in Ref. \cite{OToole2008}  and successfully compared to experiments where the GC was lifted and the axon was mechanically pulled with a cantilever. Notice however that in Ref. \cite{OToole2008}, $v$ is the microtubule velocity while, here, $v$ is the velocity of F-actin. 

\begin{figure}[!t] 
\begin{center}
\showfigures{
\includegraphics[width=0.47 \textwidth]{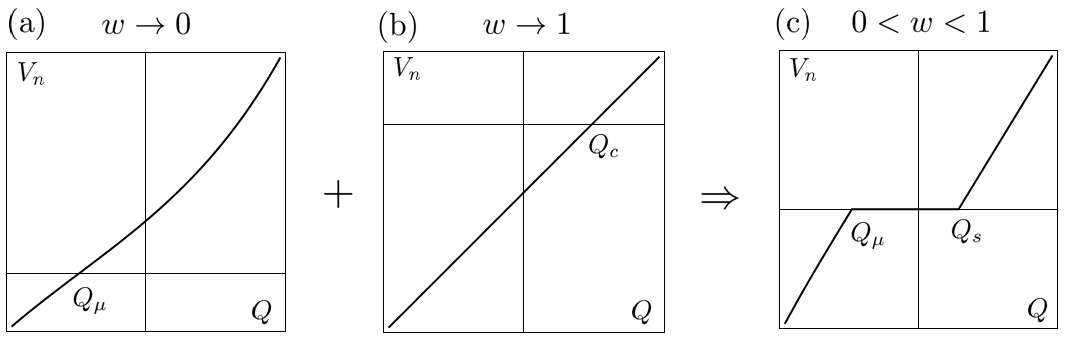}
}
\caption{
\label{fig:Scheme_VQ} (a): Velocity-force relation in the absence of a cortical acto-myosin network. (b): Velocity-force relation in the absence of a microtubule network. (c): Two thresholds Velocity-force in the general case.}
\end{center}
\end{figure}

The importance of axoplasmic contraction ($Q_{c}\not=0$) was recently demonstrated in Ref.\cite{OToole2015}. In this case, there are two possible steady states. Either  the loading is larger than the contractile stress, $Q\geq Q_{c}$, and the axon then extends at  the finite velocity $V_n=\dot{l}_n=(Q-Q_{c})/\sqrt{a}$, or  the loading is weaker than the contractile stress, $Q< Q_{c}$, and the neurite then collapses back to the soma. We sketch the force velocity $V_n(Q)$ relation in Fig.~\ref{fig:Scheme_VQ} (middle panel). This case can be referred as ``fluid-like'' growth given that the axoplasm is effectively modelled as a contractile viscous fluid. Alternatively, it was shown in Ref. \cite{molego12} that this case can also be described as a morphoelastic rod by combining an elastic response with a fast evolution of the reference configuration.\\

\paragraph*{The no-cortex case, $ \w\rightarrow 0$}
In the absence of a cortex, we combine Eq. $\text{(\ref{Model_final_sim_nd})}_1$ and Eq. $\text{(\ref{Model_final_sim_nd})}_2$, to obtain the linear equation
$$\partial_t\rho-(1+a^{-1})\partial_{xx}\rho=\epsilon(1-\rho).$$
Its long-time asymptotics, on a semi-infinite domain $x\leq l_n(t)$, can be found by considering the traveling wave reduction $y=x-l_n(t)$ in the domain $y<0$, with $l_n(t)=V_nt$. Denoting $(\ )'$ the derivative with respect to $y$, we obtain
$$-V_n\rho-(1+a^{-1})\rho''=\epsilon(1-\rho),$$ 
with boundary conditions,
$$\rho|_{0}=\text{e}^{Q-Q_{\mu}}\text{, }\partial_x\rho|_{-\infty}=0,$$ and the front velocity given by,
$$V_n=-(1+a^{-1})\text{e}^{Q_{\mu}-Q}\partial_x\rho|_{0}.$$
The solution of this linear problem is given by
$$\rho(y)= 1-\frac{p(Q)}{1+p(Q)}\text{e}^{y/l(Q)}.$$
This expression depends on $l(Q)$, which can be interpreted  as the typical size of a boundary layer, over which the chemical reaction of polymerisation/depolymerisation of microtubules is maintained out of equilibrium at the tip of the neurite:
$$l(Q)=\sqrt{\frac{(1+a^{-1})(1+p(Q))}{\epsilon}}.$$
 
The parameter $p(Q)=\text{e}^{Q-Q_{\mu}}-1>-1$ represents the driving force leading to expansion or retraction. Indeed, the tip velocity can be expressed as
$$V_n=\frac{p\sqrt{\epsilon(1+a^{-1})}}{\sqrt{1+p}}.$$
If $p<0$ ($Q<Q_{\mu}$), then the combined effect of external force and shrinking of microtubules leads to a collapse back to the soma while if $p>0$ ($Q> Q_{\mu}$) the stress provided from microtubules growth is large enough to overcome an external load and a  steady expansion is predicted.   In the absence of molecular motors, that is, without contraction, $Q_{\mu}<0$ as microtubules are able to push \cite{Dogterom1997}. But, in the presence of molecular motors,  the sign of $Q_{\mu}$ cannot be readily established \cite{OToole2015}.  We sketched the force velocity $V_n(Q)$ relation on Fig.~\ref{fig:Scheme_VQ} (left panel). In this case  the neurite is effectively modelled as a growing elastic solid.

Notice that  the velocity of the microtubules
$$v_{\mu}=(1+a^{-1})\frac{\partial_y\rho}{\rho}=\frac{p(1+a^{-1})\text{e}^{y/l}}{l(1+p-p\text{e}^{y/l})}$$ 
also displays an exponential decay away from the tip of the neurite. This behavior is consistent with the experiments of Ref. \cite{OToole2008}.

From the two limiting cases discussed above, we can obtain a general picture of the dynamics when $0<w<1$, for small and large values of the applied load $Q$: 
\begin{itemize}
\item If $Q<Q_{\mu}$, the axoplasm will collapse back to the soma. Indeed, both the cortex and the microtubule networks are in a collapse mode;
\item  if $Q>Q_{c}$,  the axon length will increase at a finite velocity given that both the cortex and the microtubules network are in extension.
\end{itemize}
Next, we consider the interval $Q\in [Q_{\mu},Q_{c}]$ by studying possible static states where a finite load does not lead to motion.

\subsection{Static states}

Static  states are the solutions of the following problem:
\begin{equation}\label{static_syst}
\left\lbrace \begin{array}{l}
 \w\partial_{xx}v-av=(1- \w)\frac{\partial_x\rho}{\rho}\\
\partial_x\left( \rho v-(1- \w)\partial_{x}\rho\right) =\epsilon(1-\rho),
\end{array}\right. 
\end{equation}
with the boundary conditions of Eq. \eqref{Model_final_BC_sim_nd} and where $l_n$ is a constant given by the condition
$$v\vert_{l_n}=(1- \w)\frac{\partial_x\rho}{\rho}\vert_{l_n(t)}.$$
While there is no obvious solution to this nonlinear problem, the following two limiting cases shed light on the general case.
\paragraph*{Large microtubule network viscosity, $\epsilon\rightarrow 0$} In the limit where the microtubule network viscosity is much larger than the acto-myosin network viscosity, we can take the limit $\epsilon\to 0$.
In this case Eq. $\text{(\ref{static_syst})}_2$ can be solved exactly:
\begin{equation}\label{static_density}
\rho(x)=\exp\left( Q_{\mu}-Q+\frac{\sigma(x)-Q}{a(1- \w)}\right),
\end{equation}
which can then be substituted into Eq. $\text{(\ref{static_syst})}_1$ to obtain,
$$ \w \partial_{xx}v-(1+a)v=0.$$
The solution of this last equation is simply
$$v(x)=(Q-Q_{c})\sqrt{\frac{ \w}{1+a}}\frac{\sinh\left(\frac{x}{\sqrt{\frac{ \w }{1+a}}}\right)}{\cosh\left(\frac{l_n}{\sqrt{\frac{ \w }{1+a}}}\right)}.$$
The stress is obtained by integrating  $v$:
$$\sigma(x)=\frac{ \w a}{1+a}(Q-Q_{c})\frac{\cosh\left(\frac{l_n}{\sqrt{\frac{ \w }{1+a}}}\right)-\cosh\left(\frac{x}{\sqrt{\frac{ \w }{1+a}}}\right)}{\cosh\left(\frac{l_n}{\sqrt{\frac{ \w }{1+a}}}\right)},$$
which can be substituted back into Eq. \eqref{static_density} to obtain a closed expression for the density. The last constraint is provided by integrating Eq.~$\text{(\ref{Model_final_sim_nd})}_2$ and requiring that for steady states the average density of microtubules is conserved, i.e., 
$$\frac{1}{l_n}\int_0^{l_n}\rho(x)dx=1.$$ 
This constraint is now an integral equation for the static length $l_n$:
\begin{eqnarray}\label{eq_integ_length}
\nonumber&&l_n\text{e}^{Q-Q_{\mu}}=\\
\nonumber&&\ \ {\Large{\int}}_0^{l_n}\exp\left\lbrace  -f_0(Q-Q_{c})\left[1- \frac{\cosh\left(\frac{x}{\sqrt{\frac{ \w }{1+a}}}\right)}{\cosh\left(\frac{l_n}{\sqrt{\frac{ \w }{1+a}}}\right)}\right] \right\rbrace  dx\\ 
\end{eqnarray}
where the constant $f_0$ reads $f_0= \w/[(1+a)(1- \w)]$. In Appendix \ref{Appen:existence}, we show that there exists a steady solution for  $Q\in [Q_{\mu},Q_s^0]$ with
\begin{equation}\label{eq_thresh_static}
Q_s^0=\frac{Q_{\mu}+f_0Q_{c}}{1+f_0}.
\end{equation}
The parameter $Q_s^0$ is an average of $Q_{\mu}$ and $Q_{c}$ weighted by the cortical width and the friction coefficients.
\begin{figure}[!t] 
\begin{center}
\showfigures{
\includegraphics[width=0.46\textwidth]{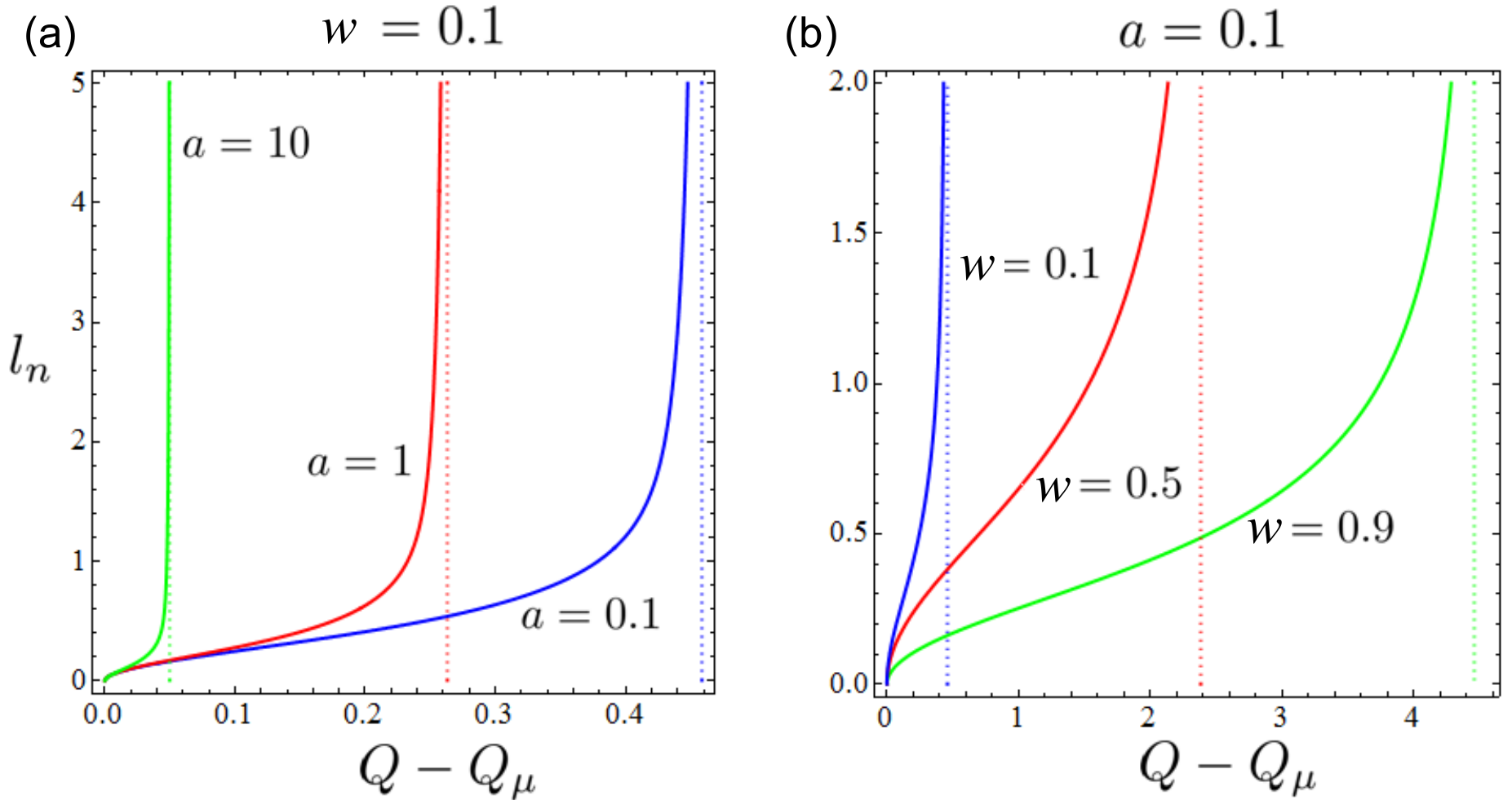}
}
\caption{
\label{fig:lengthstatic}  Length-force relations for the static solutions given by Eq. \eqref{eq_integ_length} for different values of $a$ (Fig \ref{fig:lengthstatic} (a)) and $ \w$ (Fig \ref{fig:lengthstatic} (b)). Parameter $\Delta Q_{\mu}^c=5$. Colors online }
\end{center}
\end{figure}
The numerical solution of Eq. \eqref{eq_integ_length}  between the two threshold loads is given in Fig.~\ref{fig:lengthstatic}. The static length increases monotonically from $0$ to $\infty$ between $Q_{\mu}$ and $Q_s^0$.  
  
\paragraph*{Large cortex viscosity, $\epsilon\rightarrow \infty$}
In this limit, it is clear from the right hand-side of Eq. $\text{(\ref{Model_final_sim_nd})}_2$ that $\rho$ converges to $1$ almost everywhere in the layer, except in a boundary layer close to the tip where it has to satisfy the  constraint $\rho\vert_{l_n}=\text{e}^{Q_{\mu}-Q}$. To obtain the dynamics of the moving front $l_n$ we use the piecewise linear ansatz:
$$\rho_{\epsilon}=\left\lbrace \begin{array}{l} 1\text{ if } x<l_n-\frac{1}{\epsilon}\\
\epsilon\left( \text{e}^{Q_{\mu}-Q}-1\right) (x+\frac{1}{\epsilon}-l_n)+1\text{ if } x>l_n-\frac{1}{\epsilon}.
\end{array}\right.  
$$
Using this ansatz, we solve Eq. $\text{(\ref{Model_final_sim_nd})}_1$ to obtain $v(l_n)$ asymptotically in $1/\epsilon$. To leading order, we have,
$$v(l_n)= \frac{(1- \w)(1-\text{e}^{Q_{\mu}-Q})+ \w (Q-Q_{c})}{\text{e}^{Q_{\mu}-Q}\sqrt{ \w a}}\text{th} \left(\frac{\sqrt{a}l_n}{\sqrt{ \w }}\right).$$
This value is finite because the left handside of Eq. $\text{(\ref{Model_final_sim_nd})}_1$ is a regularising elliptic operator. To leading order, the front dynamic is then given by
$$V=\dot{l}_n\sim(1- \w)\sqrt{a}\epsilon\frac{1-\text{e}^{Q_{\mu}-Q}}{\text{e}^{Q_{\mu}-Q}}.$$
We conclude that in the large $\epsilon$ regime, there is no static front unless $Q=Q_{\mu}$. For $Q>Q_{\mu}$ the axon increases indefinitely and collapses for $Q<Q_{\mu}$.\\
%

\subsection{General behavior}\label{neurite_general}

\begin{figure}[!t] 
\begin{center}
\showfigures{
\includegraphics[width=0.47\textwidth]{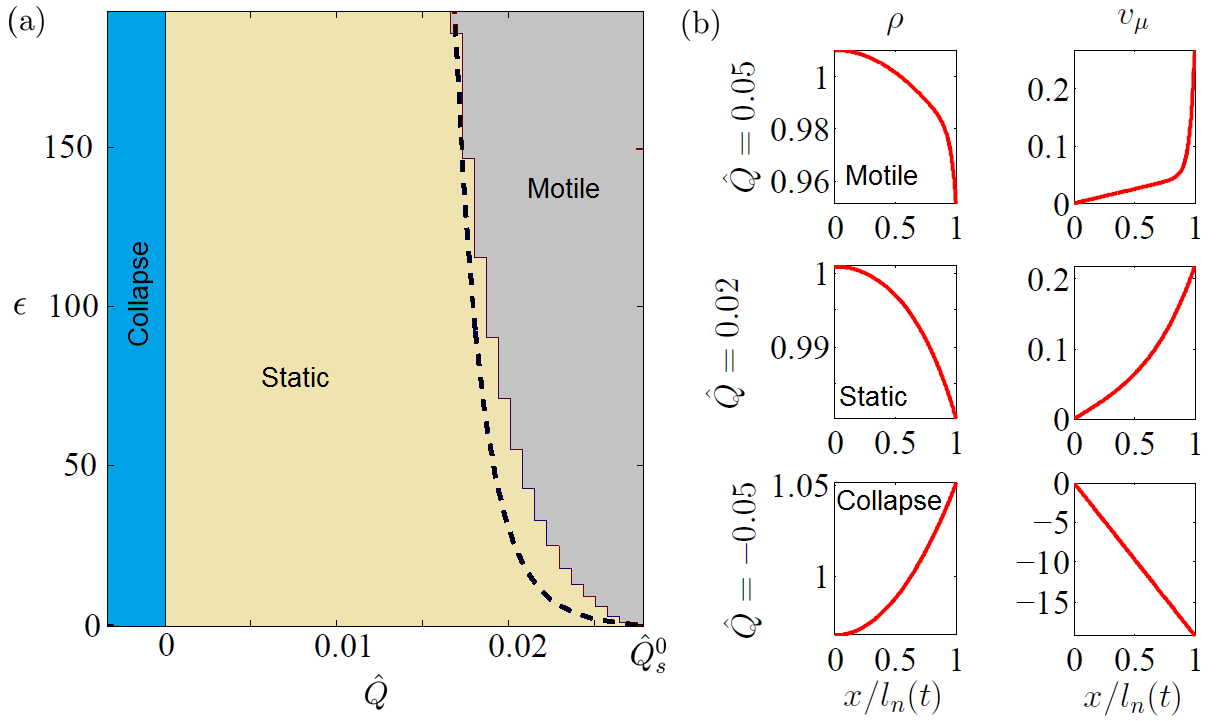}
}
\caption{
\label{fig:phasediag} (a): Numerically constructed phase diagram of the axoplasm. Parameters are $a=0.1$,  $w=0.1$ and $\Delta Q_{\mu}^c=0.3$. The dashed line is the analytic approximation from the meta-model presented in Section \ref{neurite_general}. (b): Typical steady state profiles of the microtubules velocity and density in the three phases for $\epsilon=0.001$. The numerical method is presented in Appendix~\ref{num_method}. Colors online.}
\end{center}
\end{figure}
Rather than tackling  the difficult questions of uniqueness and stability (both local and global) of the solutions that we have given in the previous sections, we use  numerical integration of problem \eqref{Model_final_sim_nd} to build the phase diagram shown on Fig.~\ref{fig:phasediag} (see Appendix~\ref{num_method} for the method). For a given set of parameters $ \w$, $a$, $\Delta Q_{\mu}^c$, we show in the $(Q ,\epsilon)$-plane the domain of existence of the three observed behaviours: Collapse, Static, and Motile. 



Essentially, the overall behaviour at finite $\epsilon$ is as follows:
\begin{itemize}
\item For  $Q<Q_{\mu}$, we observe a collapse of the neurite back to the soma. This collapse is associated with a backward  flow of microtubules along the entire axoplasm (see Fig.~\ref{fig:phasediag}) as observed experimentally \cite{OToole2015}.  We find numerically that the time to collapse decreases with increasing $\epsilon$ since the effective resistance of the microtubule network to contraction decays.
\item For  $Q_{\mu}\leq Q\leq Q_s$, where $Q_{\mu}<Q_s<Q_s^0$: we observe a stabilisation of the neurite in a static state stemming from an interplay between the growing core and the surrounding contractile sleeve. Therefore these static states may be interpreted as a tensile tightening of two parallel active networks.
\item  For $Q>Q_s$, we observe that the neurite tip moves with a finite velocity which increases with $\epsilon$. The microtubules flow increases towards the tip and develops a boundary layer at the junction with the GC as observed experimentally  \cite{OToole2008,OToole2015}. 
\end{itemize}
Accordingly, the general qualitative picture for the velocity-force relation is given in Fig.~\ref{fig:Scheme_VQ} (c).

This double force thresholds system is in  agreement with experiments \cite{Heidemann1993}. Physically, our model reveals that the applied tip stress $Q$ (positive when the neurite is pulled and negative if it pushes against an obstacle)  must be larger than the stress created by the growing microtubule network $Q_{\mu}$ to avoid collapse but it must also be larger than the effective stalling stress $Q_s$ of the entire structure to lead to steady elongation. See Fig.~\ref{fig:Scheme_VQ} (c).  Between these thresholds, the neurite effectively behaves as a neutral solid in the sense that an increase of force leads to a global strain of the neurite which acquires a new rest length. We further speculate that the oscillations in the loading at the T-domain coming from oscillating filopodia \cite{Betz2006} can lead to the small scale stop and go motion \cite{Bard2008} experimentally observed during elongation.

\subsection{Further simplifications} 

Analytical estimates can be obtained if we further simplify the model by using the fact that  microtubule growths are localised at the tip of the axoplasm and provides an effective advection velocity of the free boundary \cite{Julicher2007, Rubinstein2009, Recho2013a}.  

We assume that the axoplasm is a mixture of  contractile acto-myosin network (with fraction $ \w$) and  microtubule network (with fraction $1- \w$). In the non-dimensional notations used previously, the mechanics of the contractile phase is then given by
\begin{equation}\label{contra_phase}
\left\lbrace \begin{array}{l}
\sigma_c=\partial_xv_c+Q_c\\
\partial_x\sigma_c=\zeta_{\text{eff}}v_c\\
\partial_x\sigma_c\vert_{0}=0 \text{ and }\sigma_c\vert_{l_n}=Q.
\end{array}\right. 
\end{equation}
Similarly, the growing microtubule phase is given by
\begin{equation}\label{expan_phase}
\left\lbrace \begin{array}{l}
\sigma_{\mu}=\eta_{\text{eff}}\partial_xv_{\mu}+Q_{\mu}\\
\partial_x\sigma_{\mu}= v_{\mu}\\
\partial_x\sigma_{\mu}\vert_{0}=0 \text{ and }\sigma_{\mu}\vert_{l_n}=Q,
\end{array}\right. 
\end{equation}
where  $\zeta_{\text{eff}}=(1+a)^n$ and $\eta_{\text{eff}}=(1+\epsilon)^{-m}$ with parameters $n>0$ and $m>0$ chosen below.
The front dynamic is given by the no-flux boundary condition,
$$\dot{l}_n= \w v_c\vert_{l_n}+(1- \w)v_{\mu}\vert_{l_n}.$$
Solving Eq. \eqref{contra_phase} and \eqref{expan_phase} and assuming that $\sqrt{\eta_{\text{eff}}}$ is small enough (localised tip growth assumption) compared to $l_n$, the moving front dynamic is given by
\begin{equation}\label{simplified_front}
\dot{l}_n= \w (Q-Q_{c})\tanh(\sqrt{\zeta_{\text{eff}}}l_n)+(1- \w)\frac{Q-Q_{\mu}}{\sqrt{\eta_{\text{eff}}}}
\end{equation}
which has a similar behaviour as the full model. Namely,
\begin{itemize}
\item For $Q<Q_{\mu}$,  $l_n\rightarrow 0$ and the neurite collapses to the soma.
\item For $Q_{\mu}<Q<Q_s$, we have $l_n\rightarrow l_n^s$ and the neurite reaches a finite length given by
\begin{equation}\label{finite_length}
 l_n^s=\frac{1}{\sqrt{\zeta_{\text{eff}}}}\text{arctanh}\left(\frac{Q-Q_{\mu}}{ f_{\text{eff}} (Q_{c}-Q)}\right)
\end{equation}
where $f_{\text{eff}}= \w/(1- \w)\sqrt{\eta_{\text{eff}}/\zeta_{\text{eff}}}$.
The threshold load is given explicitly by
$$Q_s=\frac{Q_{\mu}+ f_{\text{eff}}Q_{c}}{1+ f_{\text{eff}}}.$$
To recover the value $Q_s\rightarrow Q_s^0$ (see Eq. \eqref{eq_thresh_static}) in the limit $\epsilon\rightarrow 0$, we choose $n=2$. Then, we also have $Q_s\rightarrow Q_{\mu}$ in the limit $\epsilon\rightarrow \infty$. Finally, the parameter $m\sim 1/5$ is chosen heuristically to approximate the value of $Q_s$ given by  the general phase diagram (see Fig.~\ref{fig:phasediag}).   
\item For $Q>Q_s$, we have $l_n\rightarrow \infty$ and the neurite reaches a finite velocity $V_n$ given by
\begin{equation}\label{velocity_neurite_simplified}
V_n=\frac{(1- \w)}{\sqrt{\eta_{\text{eff}}}}(1+f_{\text{eff}})\left( Q-Q_s\right) .
\end{equation}
\end{itemize}

\section{Growth cone propulsion}\label{gc_propulsion}

Ahead of the axoplasm,  we do not distinguish the lamellipodial and filopodial phases of the GC in the following 1D model. The GC is continuous with the acto-myosin cortex and we shall therefore model it as an  visco-contractile material. We use the index $gc$ to denote variables related to the GC. 

\paragraph*{Balance of mass}
The mass balance for actin reads
$$\partial_t\rho_{gc}+\partial_x(\rho_{gc} v_{gc})=-k_d\rho_{gc}$$
where $\rho_{gc}$ is the density of F-actin and $v_{gc}$, its velocity in the lab reference frame. Here, $k_d$ is the bulk depolymerisation rate \cite{Kruse2005}. 
This equation is supplemented with the kinetic boundary condition 
$$\dot{l}_{gc}=v(l_{gc}(t),t)+v_p,$$
where $v_p$ is the localised polymerisation (G-actin to F-actin) velocity at the tip of the GC \cite{Betz2006,Julicher2007, Rubinstein2009}. 
More generally, $v_p$ may also depend on the microtubules extending into the filopodium \cite{Rauch2013} and  on the external loading \cite{Julicher2007} at the proximal tip of the GC, but we shall not consider this dependence. Therefore, we assume here a stress free leading edge. As in the previous discussion, the  high compressibility of F-actin leads to the decoupling of the actin density with the front dynamic. 

\paragraph*{Balance of momentum.} In the viscous regimes, the balance of linear momentum reads
\begin{equation}\label{Force_bal_gc}
\partial_x\sigma_{gc}=\zeta_c v_{gc},
\end{equation}
where $\zeta_c$ is the friction coefficient with respect to the substrate.
 This equation is supplemented with stress boundary conditions,
\begin{equation}\label{bc_gc}
\sigma_{gc}|_{l_n(t)}=Q \text{ and } \sigma_{gc}|_{l_{gc}(t)}=0
\end{equation}
where, as before, $Q$ denotes the common traction force in the T-domain at the axoplasm/GC interface. The leading edge of the cone is assumed to be stress free. 
\paragraph*{Constitutive relation}
The constitutive relation includes both a viscous and a contractile term:
\begin{equation}\label{behav_law_gc}
\sigma_{gc}=\eta \partial_xv_{gc}+\chi \bar{c}-p.
\end{equation}
The pressure $p$ is defined numerically as a constant Lagrange multiplier associated with the conservation of the  one-dimensional volume of the GC,
\begin{equation}\label{fixed_length}
l_{gc}(t)-l_{n}(t)=L.
\end{equation} 
This constraint follows from both osmotic effects \cite{Jiang2013, Hui2014} and the fact that few compressible microtubules are engaged into filopodia  \cite{Recho2014}. 

More general models taking into account global compressibility of the GC may be required to access variation of $L$ when some rheological parameters such as the tip growth velocity or the contractility for instance are affected by drug treatments \cite{Sayyad2015}. However, our goal here is to  describe the entire neurite and  we will not discuss these finer effects further.

\paragraph*{Crawling velocity of the cone}
Combining Eq. \eqref{Force_bal_gc} and \eqref{behav_law_gc}, we obtain,
\begin{equation}\label{stress_gc}
-l_{c}^2\partial_{xx}\sigma_{gc}+\sigma_{gc}=\chi \bar{c}-p.
\end{equation}
Solving Eq. \eqref{stress_gc} with boundary conditions (see Eq. \eqref{bc_gc}) and satisfying the constraint of fixed length (see Eq.~\eqref{fixed_length}),  we obtain a closed expression for $\sigma_{gc}$ and $v_{gc}$ which we use to compute the fronts dynamic (see Ref. \cite{Recho2013a,Recho2015b} for further details):
$$\dot{l}_{gc}=\dot{l}_n=V_{gc}= \frac{1}{2}\left(v_p-\frac{Q}{\sqrt{\eta\zeta_{c}}\tanh(L/(2l_c))}\right).$$
Introducing the dimensionless velocities $\tilde V_{gc}=V_{gc}/(E/\sqrt{\eta \zeta_u})$ and $\tilde v_p=v_p/(E/\sqrt{\eta \zeta_u})$, $\tilde Q=Q/E$, this last expression becomes
$$\tilde V_{gc}= \frac{1}{2}\left(\tilde v_p-\frac{\tilde Q}{\sqrt{a}\tanh(L/(2l_c))}\right).$$
From simple physical parameters estimates (see Table \ref{t:valpar}), we have $L\gg 2l_c$, so that $\tanh(L/(2l_c))\sim 1$ and  we obtain, after dropping the tildes,  
\begin{equation}\label{velocity_cone}
V_{gc}=\frac{1}{2}\left(v_p-\frac{Q}{\sqrt{a}}\right).
\end{equation}

The GC is propelled by the polymerisation of the actin network at the leading edge which pushes the membrane forward. But, the GC is also pulled by the traction force at the interface with the axoplasm. This traction force decreases the velocity of migration if $Q>0$. The GC stops moving when $Q$ reaches the stall force
$$Q_{gc}=\sqrt{a}v_p.$$

\section{Full neurite crawling}\label{full_neurite} 

\subsection{Overall behaviour}

We can now combine  the models for the motion of the GC and the axoplasm parts to obtain a full picture of the neurite dynamics. We use the analytic relations derived in Sections \ref{neurite_general} and \ref{gc_propulsion} as they capture the main effects. We can distinguish three cases depending on the acto-myosin of the GC $Q_{gc}$:
\begin{itemize}
\item If $Q_{gc}<Q_{\mu}$, then the axon collapses to the soma in finite time. 
\item If $Q_{\mu}<Q_{gc}<Q_s$, then the axon has a finite static length. Using the approximation given by Eq. \eqref{finite_length}, we have
\begin{equation}\label{finite_length_axon}
 l^s=\frac{1}{\sqrt{\zeta_{\text{eff}}}}\text{arctanh}\left(\frac{Q_{gc}-Q_{\mu}}{ f_{\text{eff}} (Q_{c}-Q_{gc})}\right). 
\end{equation}

\item If $Q_s<Q_{gc}$, the axon acquires a finite steady state velocity. Using the approximation given by Eq.~\eqref{velocity_neurite_simplified} and \eqref{velocity_cone}, we obtain
\begin{equation}\label{velocity_total_neurite}
V=\frac{Q_{gc}-Q_s}{2\sqrt{a}+\frac{(1- \w)}{\sqrt{\eta_{\text{eff}}}}(1+f_{\text{eff}})}. 
\end{equation}
\end{itemize}
We conclude that the main behaviour of the system is captured by the relative magnitude of the three stall forces of the different neurite phases: the microtubules network ($Q_{\mu}$), the entire axoplasm ($Q_s$) and the acto-myosin of the GC ($Q_{gc}$).

\subsection{Parameter estimation}

The three loads  $Q_{\mu},\ Q_s$ and $Q_{gc}$ depend on six non-dimensional parameters directly related to measurable material coefficients. Based on Table \ref{t:valpar}, we have
\begin{table}
\scriptsize
\begin{tabular}{lll}
\hline\hline
name & symbol & typical value \\ 
\hline
F-actin viscosity & $\eta$ & $ 10^3$ Pa.s \cite{Betz2011}\\
Elasticity of microtubules & $E$& $200-400$ Pa \cite{Dennerll1989,Bernal2007,Bernal2010}\\ 
Microtubules viscosity  & $E\tau$ & $1-5\times 10^6$ Pa.s \cite{OToole2008, OToole2015}\\
Viscous friction coefficient & $\zeta_c$ &$ 10^{14}-10^{15}$ Pa.$\text{m}^{-2}\text{.s}$ \cite{OToole2008, Chan2008} \\
Contractility & $\chi \bar{c}$ &$10-10^2$ Pa \cite{Bernal2007,Bernal2010,Betz2011,OToole2015} \\
F-actin polymerisation velocity & $v_p$ &$2 \times 10^{-8}$ m.$\text{s}^{-1}$ \cite{Betz2006,Julicher2007} \\
GC Length & $L$ &$1-2\times 10^{-5}$ m \cite{Betz2011, OToole2015}\\
Cortex to axon width  & $ \w$ &$0.1$  \cite{Xu2013}\\
Friction cortex/microtubules & $\zeta_{\mu}$ &$10 \zeta_c$  (estimated) \\
hydrodynamic length $l_c$ & $\sqrt{\eta/\zeta_{c}}$ &$\sim 1.5\times 10^{-6}$ m  \\
characteristic length & $\sqrt{\eta/\zeta_{\mu}}$ &$\sim 4.4\times 10^{-7}$ m  \\
characteristic time & $\eta/E$ &  $\sim 3$ s \\
characteristic velocity & $E/\sqrt{\eta \zeta_{\mu}}$ & $\sim 1.5\times 10^{-7}$ $ \text{m.}\text{s}^{-1}$ \\
characteristic stress & $E$ & $\sim 300$ Pa \\
\hline\hline
\end{tabular}
\caption{\small Estimates of material coefficients.}
\label{t:valpar}
\end{table}
$$a\sim 0.1\text{, }\epsilon\sim 0.001\text{, } \w\sim 0.1 \text{, }v_p\sim 0.14\text{ and }Q_c\sim 0.3.$$

It is more difficult to asses the value of the microtubule network stall force $Q_{\mu}$ as the presence of molecular motors may induce contraction \cite{OToole2015}. In Ref. \cite{Rauch2013}, it is shown that the growth of microtubules engaging in filopodia can lead to a pushing stress of $-90$ Pa at the tip. However  in agreement with Ref. \cite{OToole2015}, we assume that the axonal microtubules exert a small pulling stress. Here, we choose $9$ Pa by setting the neurite velocity to about $10\mu\text{m}.\text{h}^{-1}$ leading to
$$Q_{\mu}\sim 0.03.$$

\subsection{Comparison with experiments}

Now that our model has been validated against classical pulling experiments, it is interesting to see how its predictions compare with various pharmacological tests affecting the F-actin and microtubules meshworks.\\
The effects of some classical drug treatments on the model parameters are collected in Table \ref{t:drugtreatments}.
\begin{table}
\scriptsize
\begin{tabular}{lll}
\hline\hline
Drug & effect & parameters trend \\ 
\hline
blebbistatin & inhibit myosin II contractility & $Q_c\downarrow$  \\
BDM & inhibit Myosin II contractility& $Q_c\downarrow$  \\ 
cytochalasin  & inhibit actin polymerisation& $v_p\downarrow$ ($Q_c\downarrow$ high concentration)  \\
latrunculin  & destroys the actin meshwork &$v_p\downarrow, Q_c\downarrow  $ \\
nocodazole & depolymerises microtubules & $Q_{\mu}\uparrow$ \\
epothilone B & polymerises microtubules & $Q_{\mu}\downarrow$ \\
taxol & stabilises microtubules &$\epsilon \downarrow$ ($Q_{\mu}\downarrow$ low concentration)  \\
trypsin & detaches the neurite  & $a\downarrow$\\
\hline\hline
\end{tabular}
\caption{\small Effects of classical drugs.}
\label{t:drugtreatments}
\end{table}
 
\begin{figure}[!t] 
\begin{center}
\showfigures{
\includegraphics[width=0.4\textwidth]{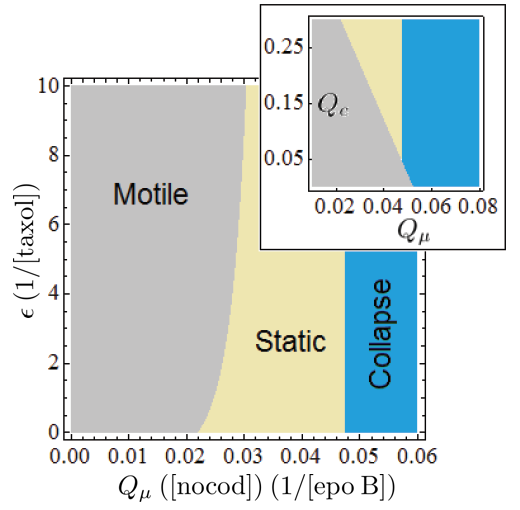}
}
\caption{
\label{fig:phasedrugs1}Perturbation of the microtubules network properties by nocodazole, taxol (inset: latrunculin and BDM reducing $Q_c$): phase diagrams of the whole neurite state using expressions of the three driving loads $Q_{\mu}$, $Q_s$ and $Q_{gc}$ (default parameters are $a\sim 0.1\text{, }\epsilon\sim 0.001\text{, } \w\sim 0.1 \text{, }v_p\sim 0.14\text{, }Q_c\sim 0.3\text{ and } Q_{\mu}\sim 0.03$). Colors online }
\end{center}
\end{figure}

\paragraph*{Retraction under microtubules depletion}
Experiments have shown that the depolymerisation of microtubules with nocodazole stops or even leads to the collapse of neurites depending on the concentration \cite{Dennerll1988, Ahmad2000, Sayyad2015}.  This treatment can be interpreted in our model as an increase of $k_d^{\mu}$ and thus an increase of $Q_{\mu}$ and is qualitatively captured in Fig.~\ref{fig:phasedrugs1}.

Physically, the depolymerisation of microtubules lowers the resistance of this growing network to contractile acto-myosin stress. As we show in the inset of Fig.~\ref{fig:phasedrugs1} and as experimentally confirmed in Ref. \cite{Ahmad2000}, nocodazole induced collapse can be rescued by a latrunculin (destroys the acto-myosin cortex) or BDM (inhibits myosin II contractile activity) which effectively reduces $Q_c$. 

Conversely, initiation of motility due to blebbistatin treatment (contractility inhibitor) is abolished if it is followed by a nocodazole treatment \cite{Hur2011}. Additionally, an increase of microtubules polymerisation using epothilone B points towards a very promising therapeutic route to promote \emph{in vivo} axonal outgrowth after injury of the spinal cord through the inhibitory environment due to the tissue scar \cite{ruschel2015systemic}. Treatment with  high concentration of taxol stabilises microtubules and slows down elongation \cite{Letourneau1984, Bamburg1986, witte2008microtubule}. This can be interpreted in the model as a decrease of $\epsilon$ and is also correctly captured as seen in Fig.~\ref{fig:phasedrugs1}. However, the effect of low concentration of taxol does not block the microtubules dynamic completely \cite{witte2008microtubule} but primarily lowers $Q_{\mu}$ (by lowering $k_d^{\mu}$), thus leading to an increase of axonal outgrowth \cite{witte2008microtubule, hellal2011microtubule}. \\

\begin{figure}[!t] 
\begin{center}
\showfigures{
\includegraphics[width=0.35\textwidth]{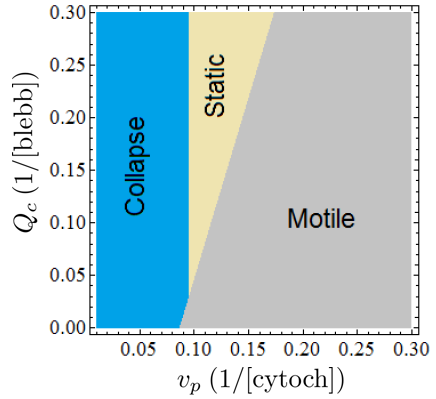}
}
\caption{
\label{fig:phasedrugs2}Perturbations of the acto-myosin meshwork by cytochalasin and blebbistatin: phase diagram of the whole neurite state using expressions of the three driving loads $Q_{\mu}$, $Q_s$ and $Q_{gc}$ (default parameters are $a\sim 0.1\text{, }\epsilon\sim 0.001\text{, } \w\sim 0.1 \text{, }v_p\sim 0.14\text{, }Q_c\sim 0.3\text{ and } Q_{\mu}\sim 0.03$). Colors online }
\end{center}
\end{figure}

\paragraph*{Treatment of the  acto-myosin meshwork}
We now turn to the treatments affecting the acto-myosin meshwork (Fig.~\ref{fig:phasedrugs2}) which has two antagonistic roles \cite{Letourneau1987, Aeschlimann2000}. On the one hand, it is pulling the axoplasm thanks to F-actin front polymerisation ($v_p$) but the contractile acto-myosin cortex is also pulling the neurite backward. Remarkably, treatment with a low concentration of cytochalasin \cite{Zheng1996} reduces only the front F-actin protrusion (no filopodia) and can be interpreted as lowering $v_p$ effectively reducing the neurite velocity. Larger concentrations, on the contrary, destroy the whole F-actin meshwork which strongly impacts the cortical contractility ($Q_c$) and leads to an increase of neurite velocity as captured by the model \cite{Dennerll1988}. More focussed experiments inhibiting contractility with blebbistatin \cite{Hur2011,Yu2012} confirm that contractility impairment robustly initiates neurite motility. Note again that cytochalasin ($v_p$ decrease)  abolishes this blebbistatin induced motility \cite{Hur2011} as the model also predicts, see Fig.~\ref{fig:phasedrugs2}.  \\

\begin{figure}[!t] 
\begin{center}
\showfigures{
\includegraphics[width=0.45\textwidth]{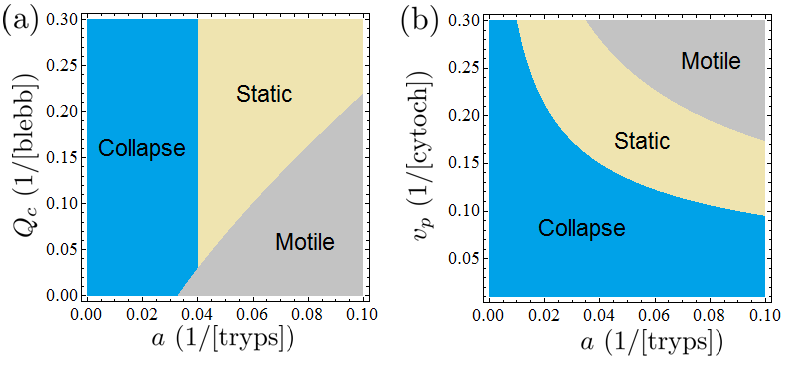}
}
\caption{
\label{fig:phaseadhe} Perturbations of the substrate adhesion by trypsin, blebbistatin and cytochalasin: phase diagrams of the whole neurite state using expressions of the three driving loads $Q_{\mu}$, $Q_s$ and $Q_{gc}$ (default parameters are $a\sim 0.1\text{, }\epsilon\sim 0.001\text{, } \w\sim 0.1 \text{, }v_p\sim 0.14\text{, }Q_c\sim 0.3\text{ and } Q_{\mu}\sim 0.03$). Colors online }
\end{center}
\end{figure}

\paragraph*{Treatment of the substrate to modify adhesions}
Adhesion of the neurite with the substrate can also be strongly reduced with trypsin, which leads to a collapse or a stall of the neurite \cite{OToole2015}. Such treatment can be modelled by lowering $a$. In Fig.~\ref{fig:phaseadhe} (a), we show that this effect is correctly captured by our model. It is also known that the motility promoting effect of myosin II inhibition is adhesiveness dependent \cite{Hur2011}. While blebbistatin promotes motility on polylysine substrates, it  lowers motility on less adherent laminin substrates \cite{Ketschek2007}. We can speculate that this is due to myosin II being strongly involved in the creation of focal adhesions for laminin substrates \cite{Brown2003, Ketschek2007}. As a result, in this case, a blebbsitatin treatment also considerably lowers adhesion ($a$) thus potentially leading to arrest (see Fig.~\ref{fig:phaseadhe} (a)). Finally, we also show in Fig.~\ref{fig:phaseadhe} (b) the effect of a cytochalasin treatment depending on the substrate adhesivity. While low level of cyotchalasin reduces the neurite velocity, we expect this effect to be attenuated on more adhesive substrates.

\section{Conclusion}
Starting from basic conservation laws, we have developed and analysed a one-dimensional mechanical  model of neurite motility based on a three-compartment cytoskeletal structure. The model supports three robust states: Collapse, Static and Motile. Collapse arises when the growth of the microtubles and the GC induced traction cannot overcome the cortical acto-myosin contractility. On the contrary, extension at a finite velocity is provoked by the GC F-actin frontal polymerisation which generates a tension promoting growth of the microtubule network and overcoming cortical contractility. Interestingly,  between these two states, the neurite can also remain static as a result of a tensile tightening between the microtubules growing network and the contractile actomyosin sleeve operating in parallel.

The respective position of the three stall forces of the microtubules, the axoplasm and the GC can be used to predict the state of the neurite and we explicitly relate these loads to measurable material parameters. This framework allows for a number of model predictions in remarkable agreement with experimental drug treatments. It is our hope that the model will be used as a guideline to design focussed experiments to discriminate the respective role of active (contractility, growth) and passive (elasticity, viscosity, substrate adhesiveness) effects impacting neurite motility and leading to a better understanding of the neuronal regeneration after a trauma.

We did not investigate the shape of neurites which is also known to be an important signature of trauma \cite{Pullarkat2006, Ahmadzadeh2014} as neurons swell or bead in response to fast pulling. To deal with this  complex problem, a two-dimensional model must be used and the osmotic pressure regulation between the inside and the outside of the neurite must be taken into account \cite{Pullarkat2006}. More generally, coupling of the cytoskeletal mechanics with the ions trafficking through channels and pumps at the plasmic membrane is an important challenge that will lead to a better insight on neurite guidance by chemical gradients as well as swelling of neurons during injury. 

\paragraph*{Acknowledgements} The authors thank Kristian Franze for helpful discussions during the international winter school on the physics of the brain held in Les Houches. P.R. acknowledges funding from OCCAM. P.R. and A.J. also acknowledge funding from the European Research Council under the European
Union's Seventh Framework Programme (FP7 2007-2013)/ERC Grant Agreement No. 306587.

\appendix

\section{Existence of solutions to Eq. (\ref{eq_integ_length})}\label{Appen:existence}

Eq. \eqref{eq_integ_length} can be rewritten in the following form,
$$e^{\hat{Q}}=\psi(l_n) \text{ where } \psi(l_n)=\int_0^1\text{e}^{[(1+f_0)Q_s^0-f_0\hat{Q}]g(u,l_n)}du,$$
with
$$g(u,l_n)=1-\frac{\cosh\left(\frac{u l_n}{l_{ \w}} \right) }{\cosh\left(\frac{l_n}{l_{ \w}} \right)}$$
and $l_{ \w}=\sqrt{\frac{ \w a}{1+a}}$.
Using the dominated convergence theorem, it follows that  
$$\psi(0)=1 \text{ and }\psi(\infty)=\text{e}^{[(1+f_0)Q_s^0-f_0\hat{Q}]}.$$
We also have  
\begin{multline*}
\frac{d\psi}{dl_n}=\frac{(1+f_0)Q_s^0-f_0\hat{Q}}{l_{ \w}\cosh\left(\frac{l_n}{l_{ \w}} \right)^2}\int_0^1 \!\!du\,\text{e}^{[(1+f_0)Q_s^0-f_0\hat{Q}]g(u,l_n)}\\
\underset{\geq \sinh\left(\frac{l_n (1-u)}{l_{ \w}} \right)\geq 0}{\underbrace{\left[ \cosh\left(\frac{ul_n}{l_{ \w}} \right)\sinh\left(\frac{l_n}{l_{ \w}} \right)
-u\cosh\left(\frac{l_n}{l_{ \w}} \right)\sinh\left(\frac{ul_n}{l_{ \w}} \right)\right]}}
.
\end{multline*}
As a result, if $\hat{Q}<Q_s$, then $\psi(l_{n})$ is an increasing function. If we additionally have $\hat{Q}>0$ then $\text{e}^{\hat{Q}}$ is strictly between $\psi(0)$ and $\psi(\infty)$ leading to the existence of a single solution of Eq. \eqref{eq_integ_length}.

\section{Numerical method} \label{num_method}

To solve the Cauchy problem of Eq. \eqref{Model_final_sim_nd}-\eqref{Model_final_BC_sim_nd}, we use the scaled space coordinate to deal with the moving boundary
\begin{equation}
  \label{eq:def:y}
  y = \frac{x}{l_n(t)},
\end{equation}
and denote the new unknown functions 
$\hat{v}(y,t)=v[l_n(t)y,t]$ and 
$\hat{\rho}(y,t)=l_n(t) \, \rho[l_n(t)y,t]$.
Eq.~$\text{(\ref{Model_final_sim_nd})}_2$ becomes
\begin{equation}
  \label{eq:evol:matter}
\partial_t\hat{\rho}
 +\frac{1}{l_n}  \partial_y\left(\hat{\rho}(\hat{v}-y\dot{l}_n)-\frac{1- \w}{l_n}\partial_{y}\hat{\rho}\right)  =l_n\epsilon\left(1-\frac{\hat{\rho}}{l_n} \right) 
\end{equation}
where the velocity field can be expressed through Eq.~$\text{(\ref{Model_final_sim_nd})}_1$ as
\begin{equation}
  \label{eq:evol:momentum}
 \frac{\w}{l_n^2}\partial_{yy}\hat{v}-a\hat{v}=\frac{1- \w}{l_n}\frac{\partial_y\hat{\rho}}{\hat{\rho}}.
\end{equation}
Accordingly, the boundary conditions of Eq. \eqref{Model_final_BC_sim_nd}
become
\begin{eqnarray}
  \label{eq:BC:evol:0}
\hat{v}\vert_0 &=& 0 \text{ and }  \partial_y\hat{v}\vert_1  = l_n(Q-Q_c),
\\
  \label{eq:BC:evol:1:v}
\partial_y\hat{\rho}\vert_0 &=& 0\text{ and }  \hat{\rho}\vert_1   = l_n\text{e}^{Q_{\mu}-Q},
\\
  \label{eq:BC:evol:1:sigma}
  \dot{l}_n &=&\hat{v}\vert_1-\frac{(1- \w)\partial_y\hat{\rho}}{l_n\hat{\rho}}\vert_1 .
\end{eqnarray}
To fully specify the system, we impose the initial conditions
$$l_n(0)=l_n^0\text{ and } \hat{\rho}(y,0)=\hat{\rho}^0(y).$$
Numerically, we did not find that the steady state phase reported in Fig. \ref{fig:phasediag} was sensitive to the choice of initial conditions.

The numerical scheme used to solve the Cauchy problem 
Eq.~(\ref{eq:evol:matter}-\ref{eq:BC:evol:1:sigma}) is based on the finite volume method \cite{Leveque2002} which allows to   conserve mass while handling  localised states without  spurious oscillations.  Two regularly-spaced grids on the same interval $[0,1]$, denoted $Z$ and $Z_d$ for its dual, are considered in parallel.
An initial condition on $\hat{\rho}$ being given on $Z$, 
Eq. \eqref{eq:evol:momentum} is solved using the boundary conditions 
of Eq. \eqref{eq:BC:evol:0} and the effective drift term $\hat{v}-y\dot{l}_n$ 
is computed on $Z_d$ using Eq. \eqref{eq:evol:momentum}. We then apply 
an upwind finite volume scheme to Eq. \eqref{eq:evol:matter} using 
the no-flux boundary conditions of Eq. \eqref{eq:BC:evol:1:v}.
This allows the computation of the updated concentration profile 
$\hat{\rho}$ on $Z$ which gives in turn the new initial
data used for the next time step and the front dynamic through Eq. \eqref{eq:BC:evol:1:sigma}. The same procedure is then repeated. 
The time interval for each time step is
adapted so that the Courant-Friedrichs-Lewy condition is uniformly 
satisfied on $Z_d$ \cite{Leveque2002}.


\begin{thebibliography}{85}%
\makeatletter
\providecommand \@ifxundefined [1]{%
 \@ifx{#1\undefined}
}%
\providecommand \@ifnum [1]{%
 \ifnum #1\expandafter \@firstoftwo
 \else \expandafter \@secondoftwo
 \fi
}%
\providecommand \@ifx [1]{%
 \ifx #1\expandafter \@firstoftwo
 \else \expandafter \@secondoftwo
 \fi
}%
\providecommand \natexlab [1]{#1}%
\providecommand \enquote  [1]{``#1''}%
\providecommand \bibnamefont  [1]{#1}%
\providecommand \bibfnamefont [1]{#1}%
\providecommand \citenamefont [1]{#1}%
\providecommand \href@noop [0]{\@secondoftwo}%
\providecommand \href [0]{\begingroup \@sanitize@url \@href}%
\providecommand \@href[1]{\@@startlink{#1}\@@href}%
\providecommand \@@href[1]{\endgroup#1\@@endlink}%
\providecommand \@sanitize@url [0]{\catcode `\\12\catcode `\$12\catcode
  `\&12\catcode `\#12\catcode `\^12\catcode `\_12\catcode `\%12\relax}%
\providecommand \@@startlink[1]{}%
\providecommand \@@endlink[0]{}%
\providecommand \url  [0]{\begingroup\@sanitize@url \@url }%
\providecommand \@url [1]{\endgroup\@href {#1}{\urlprefix }}%
\providecommand \urlprefix  [0]{URL }%
\providecommand \Eprint [0]{\href }%
\providecommand \doibase [0]{http://dx.doi.org/}%
\providecommand \selectlanguage [0]{\@gobble}%
\providecommand \bibinfo  [0]{\@secondoftwo}%
\providecommand \bibfield  [0]{\@secondoftwo}%
\providecommand \translation [1]{[#1]}%
\providecommand \BibitemOpen [0]{}%
\providecommand \bibitemStop [0]{}%
\providecommand \bibitemNoStop [0]{.\EOS\space}%
\providecommand \EOS [0]{\spacefactor3000\relax}%
\providecommand \BibitemShut  [1]{\csname bibitem#1\endcsname}%
\let\auto@bib@innerbib\@empty
\bibitem [{\citenamefont {Cajal}\ and\ \citenamefont
  {Cajal}(1911)}]{Cajal1911}%
  \BibitemOpen
  \bibfield  {author} {\bibinfo {author} {\bibfnamefont {S.~R.}\ \bibnamefont
  {Cajal}}\ and\ \bibinfo {author} {\bibfnamefont {S.}~\bibnamefont {Cajal}},\
  }\href@noop {} {\emph {\bibinfo {title} {Histologie du Syst{\`e}me Nerveux de
  l'Homme et des Vert{\'e}br{\'e}s Maloine}}}\ (\bibinfo {year}
  {1911})\BibitemShut {NoStop}%
\bibitem [{\citenamefont {Silver}\ and\ \citenamefont
  {Miller}(2004)}]{Silver2004}%
  \BibitemOpen
  \bibfield  {author} {\bibinfo {author} {\bibfnamefont {J.}~\bibnamefont
  {Silver}}\ and\ \bibinfo {author} {\bibfnamefont {J.~H.}\ \bibnamefont
  {Miller}},\ }\href@noop {} {\bibfield  {journal} {\bibinfo  {journal} {Nature
  Reviews Neuroscience}\ }\textbf {\bibinfo {volume} {5}},\ \bibinfo {pages}
  {146} (\bibinfo {year} {2004})}\BibitemShut {NoStop}%
\bibitem [{\citenamefont {Case}\ and\ \citenamefont
  {Tessier-Lavigne}(2005)}]{Case2005}%
  \BibitemOpen
  \bibfield  {author} {\bibinfo {author} {\bibfnamefont {L.~C.}\ \bibnamefont
  {Case}}\ and\ \bibinfo {author} {\bibfnamefont {M.}~\bibnamefont
  {Tessier-Lavigne}},\ }\href@noop {} {\bibfield  {journal} {\bibinfo
  {journal} {Current biology}\ }\textbf {\bibinfo {volume} {15}},\ \bibinfo
  {pages} {R749} (\bibinfo {year} {2005})}\BibitemShut {NoStop}%
\bibitem [{\citenamefont {Deumens}\ \emph {et~al.}(2010)\citenamefont
  {Deumens}, \citenamefont {Bozkurt}, \citenamefont {Meek}, \citenamefont
  {Marcus}, \citenamefont {Joosten}, \citenamefont {Weis},\ and\ \citenamefont
  {Brook}}]{Deumens2010}%
  \BibitemOpen
  \bibfield  {author} {\bibinfo {author} {\bibfnamefont {R.}~\bibnamefont
  {Deumens}}, \bibinfo {author} {\bibfnamefont {A.}~\bibnamefont {Bozkurt}},
  \bibinfo {author} {\bibfnamefont {M.~F.}\ \bibnamefont {Meek}}, \bibinfo
  {author} {\bibfnamefont {M.~A.}\ \bibnamefont {Marcus}}, \bibinfo {author}
  {\bibfnamefont {E.~A.}\ \bibnamefont {Joosten}}, \bibinfo {author}
  {\bibfnamefont {J.}~\bibnamefont {Weis}}, \ and\ \bibinfo {author}
  {\bibfnamefont {G.~A.}\ \bibnamefont {Brook}},\ }\href@noop {} {\bibfield
  {journal} {\bibinfo  {journal} {Progress in neurobiology}\ }\textbf {\bibinfo
  {volume} {92}},\ \bibinfo {pages} {245} (\bibinfo {year} {2010})}\BibitemShut
  {NoStop}%
\bibitem [{\citenamefont {Mitchison}\ and\ \citenamefont
  {Kirschner}(1988)}]{Mitchison1988}%
  \BibitemOpen
  \bibfield  {author} {\bibinfo {author} {\bibfnamefont {T.}~\bibnamefont
  {Mitchison}}\ and\ \bibinfo {author} {\bibfnamefont {M.}~\bibnamefont
  {Kirschner}},\ }\href@noop {} {\bibfield  {journal} {\bibinfo  {journal}
  {Neuron}\ }\textbf {\bibinfo {volume} {1}},\ \bibinfo {pages} {761} (\bibinfo
  {year} {1988})}\BibitemShut {NoStop}%
\bibitem [{\citenamefont {Dent}\ and\ \citenamefont
  {Gertler}(2003)}]{Dent2003}%
  \BibitemOpen
  \bibfield  {author} {\bibinfo {author} {\bibfnamefont {E.~W.}\ \bibnamefont
  {Dent}}\ and\ \bibinfo {author} {\bibfnamefont {F.~B.}\ \bibnamefont
  {Gertler}},\ }\href@noop {} {\bibfield  {journal} {\bibinfo  {journal}
  {Neuron}\ }\textbf {\bibinfo {volume} {40}},\ \bibinfo {pages} {209}
  (\bibinfo {year} {2003})}\BibitemShut {NoStop}%
\bibitem [{\citenamefont {Franze}\ and\ \citenamefont
  {Guck}(2010)}]{Franze2010}%
  \BibitemOpen
  \bibfield  {author} {\bibinfo {author} {\bibfnamefont {K.}~\bibnamefont
  {Franze}}\ and\ \bibinfo {author} {\bibfnamefont {J.}~\bibnamefont {Guck}},\
  }\href@noop {} {\bibfield  {journal} {\bibinfo  {journal} {Reports on
  Progress in Physics}\ }\textbf {\bibinfo {volume} {73}},\ \bibinfo {pages}
  {094601} (\bibinfo {year} {2010})}\BibitemShut {NoStop}%
\bibitem [{\citenamefont {Coles}\ and\ \citenamefont
  {Bradke}(2015)}]{Coles2015}%
  \BibitemOpen
  \bibfield  {author} {\bibinfo {author} {\bibfnamefont {C.~H.}\ \bibnamefont
  {Coles}}\ and\ \bibinfo {author} {\bibfnamefont {F.}~\bibnamefont {Bradke}},\
  }\href@noop {} {\bibfield  {journal} {\bibinfo  {journal} {Current Biology}\
  }\textbf {\bibinfo {volume} {25}},\ \bibinfo {pages} {R677} (\bibinfo {year}
  {2015})}\BibitemShut {NoStop}%
\bibitem [{\citenamefont {Aeschlimann}\ and\ \citenamefont
  {Tettoni}(2001)}]{Aeschlimann2001}%
  \BibitemOpen
  \bibfield  {author} {\bibinfo {author} {\bibfnamefont {M.}~\bibnamefont
  {Aeschlimann}}\ and\ \bibinfo {author} {\bibfnamefont {L.}~\bibnamefont
  {Tettoni}},\ }\href@noop {} {\bibfield  {journal} {\bibinfo  {journal}
  {Neurocomputing}\ }\textbf {\bibinfo {volume} {38}},\ \bibinfo {pages} {87}
  (\bibinfo {year} {2001})}\BibitemShut {NoStop}%
\bibitem [{\citenamefont {Dickson}(2002)}]{Dickson2002}%
  \BibitemOpen
  \bibfield  {author} {\bibinfo {author} {\bibfnamefont {B.~J.}\ \bibnamefont
  {Dickson}},\ }\href@noop {} {\bibfield  {journal} {\bibinfo  {journal}
  {Science}\ }\textbf {\bibinfo {volume} {298}},\ \bibinfo {pages} {1959}
  (\bibinfo {year} {2002})}\BibitemShut {NoStop}%
\bibitem [{\citenamefont {Lowery}\ and\ \citenamefont
  {Van~Vactor}(2009)}]{Lowery2009}%
  \BibitemOpen
  \bibfield  {author} {\bibinfo {author} {\bibfnamefont {L.~A.}\ \bibnamefont
  {Lowery}}\ and\ \bibinfo {author} {\bibfnamefont {D.}~\bibnamefont
  {Van~Vactor}},\ }\href@noop {} {\bibfield  {journal} {\bibinfo  {journal}
  {Nature reviews Molecular cell biology}\ }\textbf {\bibinfo {volume} {10}},\
  \bibinfo {pages} {332} (\bibinfo {year} {2009})}\BibitemShut {NoStop}%
\bibitem [{\citenamefont {Reingruber}\ and\ \citenamefont
  {Holcman}(2014)}]{Reingruber2014}%
  \BibitemOpen
  \bibfield  {author} {\bibinfo {author} {\bibfnamefont {J.}~\bibnamefont
  {Reingruber}}\ and\ \bibinfo {author} {\bibfnamefont {D.}~\bibnamefont
  {Holcman}},\ }in\ \href@noop {} {\emph {\bibinfo {booktitle} {Seminars in
  cell \& developmental biology}}},\ Vol.~\bibinfo {volume} {35}\ (\bibinfo
  {organization} {Elsevier},\ \bibinfo {year} {2014})\ pp.\ \bibinfo {pages}
  {189--202}\BibitemShut {NoStop}%
\bibitem [{\citenamefont {Franze}\ \emph {et~al.}(2009)\citenamefont {Franze},
  \citenamefont {Gerdelmann}, \citenamefont {Weick}, \citenamefont {Betz},
  \citenamefont {Pawlizak}, \citenamefont {Lakadamyali}, \citenamefont {Bayer},
  \citenamefont {Rillich}, \citenamefont {G{\"o}gler}, \citenamefont {Lu} \emph
  {et~al.}}]{Franze2009}%
  \BibitemOpen
  \bibfield  {author} {\bibinfo {author} {\bibfnamefont {K.}~\bibnamefont
  {Franze}}, \bibinfo {author} {\bibfnamefont {J.}~\bibnamefont {Gerdelmann}},
  \bibinfo {author} {\bibfnamefont {M.}~\bibnamefont {Weick}}, \bibinfo
  {author} {\bibfnamefont {T.}~\bibnamefont {Betz}}, \bibinfo {author}
  {\bibfnamefont {S.}~\bibnamefont {Pawlizak}}, \bibinfo {author}
  {\bibfnamefont {M.}~\bibnamefont {Lakadamyali}}, \bibinfo {author}
  {\bibfnamefont {J.}~\bibnamefont {Bayer}}, \bibinfo {author} {\bibfnamefont
  {K.}~\bibnamefont {Rillich}}, \bibinfo {author} {\bibfnamefont
  {M.}~\bibnamefont {G{\"o}gler}}, \bibinfo {author} {\bibfnamefont {Y.-B.}\
  \bibnamefont {Lu}},  \emph {et~al.},\ }\href@noop {} {\bibfield  {journal}
  {\bibinfo  {journal} {Biophysical journal}\ }\textbf {\bibinfo {volume}
  {97}},\ \bibinfo {pages} {1883} (\bibinfo {year} {2009})}\BibitemShut
  {NoStop}%
\bibitem [{\citenamefont {Ouyang}\ \emph {et~al.}(2013)\citenamefont {Ouyang},
  \citenamefont {Nauman},\ and\ \citenamefont {Shi}}]{Ouyang2013}%
  \BibitemOpen
  \bibfield  {author} {\bibinfo {author} {\bibfnamefont {H.}~\bibnamefont
  {Ouyang}}, \bibinfo {author} {\bibfnamefont {E.}~\bibnamefont {Nauman}}, \
  and\ \bibinfo {author} {\bibfnamefont {R.}~\bibnamefont {Shi}},\ }\href@noop
  {} {\bibfield  {journal} {\bibinfo  {journal} {J Biol Eng}\ }\textbf
  {\bibinfo {volume} {7}},\ \bibinfo {pages} {21} (\bibinfo {year}
  {2013})}\BibitemShut {NoStop}%
\bibitem [{\citenamefont {Brown}\ and\ \citenamefont
  {Bridgman}(2003)}]{Brown2003}%
  \BibitemOpen
  \bibfield  {author} {\bibinfo {author} {\bibfnamefont {J.}~\bibnamefont
  {Brown}}\ and\ \bibinfo {author} {\bibfnamefont {P.~C.}\ \bibnamefont
  {Bridgman}},\ }\href@noop {} {\bibfield  {journal} {\bibinfo  {journal}
  {Journal of Histochemistry \& Cytochemistry}\ }\textbf {\bibinfo {volume}
  {51}},\ \bibinfo {pages} {421} (\bibinfo {year} {2003})}\BibitemShut
  {NoStop}%
\bibitem [{\citenamefont {Roossien}\ \emph {et~al.}(2014)\citenamefont
  {Roossien}, \citenamefont {Lamoureux},\ and\ \citenamefont
  {Miller}}]{Roossien2014}%
  \BibitemOpen
  \bibfield  {author} {\bibinfo {author} {\bibfnamefont {D.~H.}\ \bibnamefont
  {Roossien}}, \bibinfo {author} {\bibfnamefont {P.}~\bibnamefont {Lamoureux}},
  \ and\ \bibinfo {author} {\bibfnamefont {K.~E.}\ \bibnamefont {Miller}},\
  }\href@noop {} {\bibfield  {journal} {\bibinfo  {journal} {Journal of cell
  science}\ }\textbf {\bibinfo {volume} {127}},\ \bibinfo {pages} {3593}
  (\bibinfo {year} {2014})}\BibitemShut {NoStop}%
\bibitem [{\citenamefont {Lu}\ \emph {et~al.}(2013)\citenamefont {Lu},
  \citenamefont {Fox}, \citenamefont {Lakonishok}, \citenamefont {Davidson},\
  and\ \citenamefont {Gelfand}}]{Lu2013}%
  \BibitemOpen
  \bibfield  {author} {\bibinfo {author} {\bibfnamefont {W.}~\bibnamefont
  {Lu}}, \bibinfo {author} {\bibfnamefont {P.}~\bibnamefont {Fox}}, \bibinfo
  {author} {\bibfnamefont {M.}~\bibnamefont {Lakonishok}}, \bibinfo {author}
  {\bibfnamefont {M.~W.}\ \bibnamefont {Davidson}}, \ and\ \bibinfo {author}
  {\bibfnamefont {V.~I.}\ \bibnamefont {Gelfand}},\ }\href@noop {} {\bibfield
  {journal} {\bibinfo  {journal} {Current Biology}\ }\textbf {\bibinfo {volume}
  {23}},\ \bibinfo {pages} {1018} (\bibinfo {year} {2013})}\BibitemShut
  {NoStop}%
\bibitem [{\citenamefont {Ahmadzadeh}\ \emph {et~al.}(2014)\citenamefont
  {Ahmadzadeh}, \citenamefont {Smith},\ and\ \citenamefont
  {Shenoy}}]{Ahmadzadeh2014}%
  \BibitemOpen
  \bibfield  {author} {\bibinfo {author} {\bibfnamefont {H.}~\bibnamefont
  {Ahmadzadeh}}, \bibinfo {author} {\bibfnamefont {D.~H.}\ \bibnamefont
  {Smith}}, \ and\ \bibinfo {author} {\bibfnamefont {V.~B.}\ \bibnamefont
  {Shenoy}},\ }\href@noop {} {\bibfield  {journal} {\bibinfo  {journal}
  {Biophysical journal}\ }\textbf {\bibinfo {volume} {106}},\ \bibinfo {pages}
  {1123} (\bibinfo {year} {2014})}\BibitemShut {NoStop}%
\bibitem [{\citenamefont {Julicher}\ \emph {et~al.}(2007)\citenamefont
  {Julicher}, \citenamefont {Kruse}, \citenamefont {Prost},\ and\ \citenamefont
  {Joanny}}]{Julicher2007}%
  \BibitemOpen
  \bibfield  {author} {\bibinfo {author} {\bibfnamefont {F.}~\bibnamefont
  {Julicher}}, \bibinfo {author} {\bibfnamefont {K.}~\bibnamefont {Kruse}},
  \bibinfo {author} {\bibfnamefont {J.}~\bibnamefont {Prost}}, \ and\ \bibinfo
  {author} {\bibfnamefont {J.-F.}\ \bibnamefont {Joanny}},\ }\href@noop {}
  {\bibfield  {journal} {\bibinfo  {journal} {Physics Reports}\ }\textbf
  {\bibinfo {volume} {449}},\ \bibinfo {pages} {3} (\bibinfo {year}
  {2007})}\BibitemShut {NoStop}%
\bibitem [{\citenamefont {Verkhovsky}\ \emph {et~al.}(1999)\citenamefont
  {Verkhovsky}, \citenamefont {Svitkina},\ and\ \citenamefont
  {Borisy}}]{Verkhovsky1999}%
  \BibitemOpen
  \bibfield  {author} {\bibinfo {author} {\bibfnamefont {A.~B.}\ \bibnamefont
  {Verkhovsky}}, \bibinfo {author} {\bibfnamefont {T.~M.}\ \bibnamefont
  {Svitkina}}, \ and\ \bibinfo {author} {\bibfnamefont {G.~G.}\ \bibnamefont
  {Borisy}},\ }\href@noop {} {\bibfield  {journal} {\bibinfo  {journal}
  {Current Biology}\ }\textbf {\bibinfo {volume} {9}},\ \bibinfo {pages} {11}
  (\bibinfo {year} {1999})}\BibitemShut {NoStop}%
\bibitem [{\citenamefont {Medeiros}\ \emph {et~al.}(2006)\citenamefont
  {Medeiros}, \citenamefont {Burnette},\ and\ \citenamefont
  {Forscher}}]{Medeiros2006}%
  \BibitemOpen
  \bibfield  {author} {\bibinfo {author} {\bibfnamefont {N.~A.}\ \bibnamefont
  {Medeiros}}, \bibinfo {author} {\bibfnamefont {D.~T.}\ \bibnamefont
  {Burnette}}, \ and\ \bibinfo {author} {\bibfnamefont {P.}~\bibnamefont
  {Forscher}},\ }\href@noop {} {\bibfield  {journal} {\bibinfo  {journal}
  {Nature cell biology}\ }\textbf {\bibinfo {volume} {8}},\ \bibinfo {pages}
  {216} (\bibinfo {year} {2006})}\BibitemShut {NoStop}%
\bibitem [{\citenamefont {Bard}\ \emph {et~al.}(2008)\citenamefont {Bard},
  \citenamefont {Boscher}, \citenamefont {Lambert}, \citenamefont {M{\`e}ge},
  \citenamefont {Choquet},\ and\ \citenamefont {Thoumine}}]{Bard2008}%
  \BibitemOpen
  \bibfield  {author} {\bibinfo {author} {\bibfnamefont {L.}~\bibnamefont
  {Bard}}, \bibinfo {author} {\bibfnamefont {C.}~\bibnamefont {Boscher}},
  \bibinfo {author} {\bibfnamefont {M.}~\bibnamefont {Lambert}}, \bibinfo
  {author} {\bibfnamefont {R.-M.}\ \bibnamefont {M{\`e}ge}}, \bibinfo {author}
  {\bibfnamefont {D.}~\bibnamefont {Choquet}}, \ and\ \bibinfo {author}
  {\bibfnamefont {O.}~\bibnamefont {Thoumine}},\ }\href@noop {} {\bibfield
  {journal} {\bibinfo  {journal} {The Journal of Neuroscience}\ }\textbf
  {\bibinfo {volume} {28}},\ \bibinfo {pages} {5879} (\bibinfo {year}
  {2008})}\BibitemShut {NoStop}%
\bibitem [{\citenamefont {Chan}\ and\ \citenamefont {Odde}(2008)}]{Chan2008}%
  \BibitemOpen
  \bibfield  {author} {\bibinfo {author} {\bibfnamefont {C.~E.}\ \bibnamefont
  {Chan}}\ and\ \bibinfo {author} {\bibfnamefont {D.~J.}\ \bibnamefont
  {Odde}},\ }\href@noop {} {\bibfield  {journal} {\bibinfo  {journal}
  {Science}\ }\textbf {\bibinfo {volume} {322}},\ \bibinfo {pages} {1687}
  (\bibinfo {year} {2008})}\BibitemShut {NoStop}%
\bibitem [{\citenamefont {Vallee}\ \emph {et~al.}(2009)\citenamefont {Vallee},
  \citenamefont {Seale},\ and\ \citenamefont {Tsai}}]{Vallee2009}%
  \BibitemOpen
  \bibfield  {author} {\bibinfo {author} {\bibfnamefont {R.~B.}\ \bibnamefont
  {Vallee}}, \bibinfo {author} {\bibfnamefont {G.~E.}\ \bibnamefont {Seale}}, \
  and\ \bibinfo {author} {\bibfnamefont {J.-W.}\ \bibnamefont {Tsai}},\
  }\href@noop {} {\bibfield  {journal} {\bibinfo  {journal} {Trends in cell
  biology}\ }\textbf {\bibinfo {volume} {19}},\ \bibinfo {pages} {347}
  (\bibinfo {year} {2009})}\BibitemShut {NoStop}%
\bibitem [{\citenamefont {Aeschlimann}(2000)}]{Aeschlimann2000}%
  \BibitemOpen
  \bibfield  {author} {\bibinfo {author} {\bibfnamefont {M.}~\bibnamefont
  {Aeschlimann}},\ }\href@noop {} {\emph {\bibinfo {title} {Biophysical Models
  of Axonal Path Finding}}}\ (\bibinfo {year} {2000})\BibitemShut {NoStop}%
\bibitem [{\citenamefont {Kiddie}\ \emph {et~al.}(2005)\citenamefont {Kiddie},
  \citenamefont {McLean}, \citenamefont {Van~Ooyen},\ and\ \citenamefont
  {Graham}}]{Kiddie2005}%
  \BibitemOpen
  \bibfield  {author} {\bibinfo {author} {\bibfnamefont {G.}~\bibnamefont
  {Kiddie}}, \bibinfo {author} {\bibfnamefont {D.}~\bibnamefont {McLean}},
  \bibinfo {author} {\bibfnamefont {A.}~\bibnamefont {Van~Ooyen}}, \ and\
  \bibinfo {author} {\bibfnamefont {B.}~\bibnamefont {Graham}},\ }\href@noop {}
  {\bibfield  {journal} {\bibinfo  {journal} {Progress in brain research}\
  }\textbf {\bibinfo {volume} {147}},\ \bibinfo {pages} {67} (\bibinfo {year}
  {2005})}\BibitemShut {NoStop}%
\bibitem [{\citenamefont {Suter}\ and\ \citenamefont
  {Miller}(2011)}]{Suter2011}%
  \BibitemOpen
  \bibfield  {author} {\bibinfo {author} {\bibfnamefont {D.~M.}\ \bibnamefont
  {Suter}}\ and\ \bibinfo {author} {\bibfnamefont {K.~E.}\ \bibnamefont
  {Miller}},\ }\href@noop {} {\bibfield  {journal} {\bibinfo  {journal}
  {Progress in neurobiology}\ }\textbf {\bibinfo {volume} {94}},\ \bibinfo
  {pages} {91} (\bibinfo {year} {2011})}\BibitemShut {NoStop}%
\bibitem [{\citenamefont {Peskin}\ \emph {et~al.}(1993)\citenamefont {Peskin},
  \citenamefont {Odell},\ and\ \citenamefont {Oster}}]{Peskin1993}%
  \BibitemOpen
  \bibfield  {author} {\bibinfo {author} {\bibfnamefont {C.~S.}\ \bibnamefont
  {Peskin}}, \bibinfo {author} {\bibfnamefont {G.~M.}\ \bibnamefont {Odell}}, \
  and\ \bibinfo {author} {\bibfnamefont {G.~F.}\ \bibnamefont {Oster}},\
  }\href@noop {} {\bibfield  {journal} {\bibinfo  {journal} {Biophysical
  journal}\ }\textbf {\bibinfo {volume} {65}},\ \bibinfo {pages} {316}
  (\bibinfo {year} {1993})}\BibitemShut {NoStop}%
\bibitem [{\citenamefont {Mogilner}\ and\ \citenamefont
  {Oster}(2003)}]{Mogilner2003}%
  \BibitemOpen
  \bibfield  {author} {\bibinfo {author} {\bibfnamefont {A.}~\bibnamefont
  {Mogilner}}\ and\ \bibinfo {author} {\bibfnamefont {G.}~\bibnamefont
  {Oster}},\ }\href@noop {} {\bibfield  {journal} {\bibinfo  {journal}
  {Biophysical journal}\ }\textbf {\bibinfo {volume} {84}},\ \bibinfo {pages}
  {1591} (\bibinfo {year} {2003})}\BibitemShut {NoStop}%
\bibitem [{\citenamefont {Bray}(1979)}]{Bray1979}%
  \BibitemOpen
  \bibfield  {author} {\bibinfo {author} {\bibfnamefont {D.}~\bibnamefont
  {Bray}},\ }\href@noop {} {\bibfield  {journal} {\bibinfo  {journal} {Journal
  of cell science}\ }\textbf {\bibinfo {volume} {37}},\ \bibinfo {pages} {391}
  (\bibinfo {year} {1979})}\BibitemShut {NoStop}%
\bibitem [{\citenamefont {Lamoureux}\ \emph {et~al.}(1989)\citenamefont
  {Lamoureux}, \citenamefont {Buxbaum},\ and\ \citenamefont
  {Heidemann}}]{Lamoureux1989}%
  \BibitemOpen
  \bibfield  {author} {\bibinfo {author} {\bibfnamefont {P.}~\bibnamefont
  {Lamoureux}}, \bibinfo {author} {\bibfnamefont {R.~E.}\ \bibnamefont
  {Buxbaum}}, \ and\ \bibinfo {author} {\bibfnamefont {S.~R.}\ \bibnamefont
  {Heidemann}},\ }\href@noop {} {\bibfield  {journal} {\bibinfo  {journal}
  {Nature}\ }\textbf {\bibinfo {volume} {340}},\ \bibinfo {pages} {159}
  (\bibinfo {year} {1989})}\BibitemShut {NoStop}%
\bibitem [{\citenamefont {O'Toole}\ \emph {et~al.}(2008)\citenamefont
  {O'Toole}, \citenamefont {Lamoureux},\ and\ \citenamefont
  {Miller}}]{OToole2008}%
  \BibitemOpen
  \bibfield  {author} {\bibinfo {author} {\bibfnamefont {M.}~\bibnamefont
  {O'Toole}}, \bibinfo {author} {\bibfnamefont {P.}~\bibnamefont {Lamoureux}},
  \ and\ \bibinfo {author} {\bibfnamefont {K.~E.}\ \bibnamefont {Miller}},\
  }\href@noop {} {\bibfield  {journal} {\bibinfo  {journal} {Biophysical
  journal}\ }\textbf {\bibinfo {volume} {94}},\ \bibinfo {pages} {2610}
  (\bibinfo {year} {2008})}\BibitemShut {NoStop}%
\bibitem [{\citenamefont {O'Toole}\ \emph {et~al.}(2015)\citenamefont
  {O'Toole}, \citenamefont {Lamoureux},\ and\ \citenamefont
  {Miller}}]{OToole2015}%
  \BibitemOpen
  \bibfield  {author} {\bibinfo {author} {\bibfnamefont {M.}~\bibnamefont
  {O'Toole}}, \bibinfo {author} {\bibfnamefont {P.}~\bibnamefont {Lamoureux}},
  \ and\ \bibinfo {author} {\bibfnamefont {K.~E.}\ \bibnamefont {Miller}},\
  }\href@noop {} {\bibfield  {journal} {\bibinfo  {journal} {Biophysical
  journal}\ }\textbf {\bibinfo {volume} {108}},\ \bibinfo {pages} {1027}
  (\bibinfo {year} {2015})}\BibitemShut {NoStop}%
\bibitem [{\citenamefont {Van~Veen}\ and\ \citenamefont
  {Van~Pelt}(1994)}]{Van1994}%
  \BibitemOpen
  \bibfield  {author} {\bibinfo {author} {\bibfnamefont {M.~P.}\ \bibnamefont
  {Van~Veen}}\ and\ \bibinfo {author} {\bibfnamefont {J.}~\bibnamefont
  {Van~Pelt}},\ }\href@noop {} {\bibfield  {journal} {\bibinfo  {journal}
  {Bulletin of mathematical biology}\ }\textbf {\bibinfo {volume} {56}},\
  \bibinfo {pages} {249} (\bibinfo {year} {1994})}\BibitemShut {NoStop}%
\bibitem [{\citenamefont {Rauch}\ \emph {et~al.}(2013)\citenamefont {Rauch},
  \citenamefont {Heine}, \citenamefont {Goettgens},\ and\ \citenamefont
  {K{\"a}s}}]{Rauch2013}%
  \BibitemOpen
  \bibfield  {author} {\bibinfo {author} {\bibfnamefont {P.}~\bibnamefont
  {Rauch}}, \bibinfo {author} {\bibfnamefont {P.}~\bibnamefont {Heine}},
  \bibinfo {author} {\bibfnamefont {B.}~\bibnamefont {Goettgens}}, \ and\
  \bibinfo {author} {\bibfnamefont {J.~A.}\ \bibnamefont {K{\"a}s}},\
  }\href@noop {} {\bibfield  {journal} {\bibinfo  {journal} {New Journal of
  Physics}\ }\textbf {\bibinfo {volume} {15}},\ \bibinfo {pages} {015007}
  (\bibinfo {year} {2013})}\BibitemShut {NoStop}%
\bibitem [{\citenamefont {Garc\'{i}a}\ \emph {et~al.}(2012)\citenamefont
  {Garc\'{i}a}, \citenamefont {Pe\~{n}a}, \citenamefont {Mchugh},\ and\
  \citenamefont {J\'{e}rusalem}}]{Garcia2012}%
  \BibitemOpen
  \bibfield  {author} {\bibinfo {author} {\bibfnamefont {J.}~\bibnamefont
  {Garc\'{i}a}}, \bibinfo {author} {\bibfnamefont {J.}~\bibnamefont
  {Pe\~{n}a}}, \bibinfo {author} {\bibfnamefont {S.}~\bibnamefont {Mchugh}}, \
  and\ \bibinfo {author} {\bibfnamefont {A.}~\bibnamefont {J\'{e}rusalem}},\
  }\href@noop {} {\bibfield  {journal} {\bibinfo  {journal} {Computer modeling
  in engineering and sciences}\ }\textbf {\bibinfo {volume} {87}},\ \bibinfo
  {pages} {411} (\bibinfo {year} {2012})}\BibitemShut {NoStop}%
\bibitem [{\citenamefont {Dogterom}\ and\ \citenamefont
  {Yurke}(1997)}]{Dogterom1997}%
  \BibitemOpen
  \bibfield  {author} {\bibinfo {author} {\bibfnamefont {M.}~\bibnamefont
  {Dogterom}}\ and\ \bibinfo {author} {\bibfnamefont {B.}~\bibnamefont
  {Yurke}},\ }\href {\doibase 10.1126/science.278.5339.856} {\bibfield
  {journal} {\bibinfo  {journal} {Science}\ }\textbf {\bibinfo {volume}
  {278}},\ \bibinfo {pages} {856} (\bibinfo {year} {1997})}\BibitemShut
  {NoStop}%
\bibitem [{\citenamefont {Buxbaum}\ and\ \citenamefont
  {Heidemann}(1992)}]{Buxbaum1992}%
  \BibitemOpen
  \bibfield  {author} {\bibinfo {author} {\bibfnamefont {R.}~\bibnamefont
  {Buxbaum}}\ and\ \bibinfo {author} {\bibfnamefont {S.}~\bibnamefont
  {Heidemann}},\ }\href@noop {} {\bibfield  {journal} {\bibinfo  {journal}
  {Journal of theoretical biology}\ }\textbf {\bibinfo {volume} {155}},\
  \bibinfo {pages} {409} (\bibinfo {year} {1992})}\BibitemShut {NoStop}%
\bibitem [{\citenamefont {Letourneau}\ \emph {et~al.}(1987)\citenamefont
  {Letourneau}, \citenamefont {Shattuck},\ and\ \citenamefont
  {Ressler}}]{Letourneau1987}%
  \BibitemOpen
  \bibfield  {author} {\bibinfo {author} {\bibfnamefont {P.~C.}\ \bibnamefont
  {Letourneau}}, \bibinfo {author} {\bibfnamefont {T.~A.}\ \bibnamefont
  {Shattuck}}, \ and\ \bibinfo {author} {\bibfnamefont {A.~H.}\ \bibnamefont
  {Ressler}},\ }\href@noop {} {\bibfield  {journal} {\bibinfo  {journal} {Cell
  motility and the cytoskeleton}\ }\textbf {\bibinfo {volume} {8}},\ \bibinfo
  {pages} {193} (\bibinfo {year} {1987})}\BibitemShut {NoStop}%
\bibitem [{\citenamefont {Dennerll}\ \emph {et~al.}(1988)\citenamefont
  {Dennerll}, \citenamefont {Joshi}, \citenamefont {Steel}, \citenamefont
  {Buxbaum},\ and\ \citenamefont {Heidemann}}]{Dennerll1988}%
  \BibitemOpen
  \bibfield  {author} {\bibinfo {author} {\bibfnamefont {T.}~\bibnamefont
  {Dennerll}}, \bibinfo {author} {\bibfnamefont {H.}~\bibnamefont {Joshi}},
  \bibinfo {author} {\bibfnamefont {V.}~\bibnamefont {Steel}}, \bibinfo
  {author} {\bibfnamefont {R.}~\bibnamefont {Buxbaum}}, \ and\ \bibinfo
  {author} {\bibfnamefont {S.}~\bibnamefont {Heidemann}},\ }\href@noop {}
  {\bibfield  {journal} {\bibinfo  {journal} {The Journal of cell biology}\
  }\textbf {\bibinfo {volume} {107}},\ \bibinfo {pages} {665} (\bibinfo {year}
  {1988})}\BibitemShut {NoStop}%
\bibitem [{\citenamefont {Kollins}\ \emph {et~al.}(2009)\citenamefont
  {Kollins}, \citenamefont {Hu}, \citenamefont {Bridgman}, \citenamefont
  {Huang},\ and\ \citenamefont {Gallo}}]{Kollins2009}%
  \BibitemOpen
  \bibfield  {author} {\bibinfo {author} {\bibfnamefont {K.}~\bibnamefont
  {Kollins}}, \bibinfo {author} {\bibfnamefont {J.}~\bibnamefont {Hu}},
  \bibinfo {author} {\bibfnamefont {P.}~\bibnamefont {Bridgman}}, \bibinfo
  {author} {\bibfnamefont {Y.-Q.}\ \bibnamefont {Huang}}, \ and\ \bibinfo
  {author} {\bibfnamefont {G.}~\bibnamefont {Gallo}},\ }\href@noop {}
  {\bibfield  {journal} {\bibinfo  {journal} {Developmental neurobiology}\
  }\textbf {\bibinfo {volume} {69}},\ \bibinfo {pages} {279} (\bibinfo {year}
  {2009})}\BibitemShut {NoStop}%
\bibitem [{\citenamefont {Hur}\ \emph {et~al.}(2011)\citenamefont {Hur},
  \citenamefont {Yang}, \citenamefont {Kim}, \citenamefont {Byun},
  \citenamefont {Xu}, \citenamefont {Nicovich}, \citenamefont {Cheong},
  \citenamefont {Levchenko}, \citenamefont {Thakor}, \citenamefont {Zhou} \emph
  {et~al.}}]{Hur2011}%
  \BibitemOpen
  \bibfield  {author} {\bibinfo {author} {\bibfnamefont {E.-M.}\ \bibnamefont
  {Hur}}, \bibinfo {author} {\bibfnamefont {I.~H.}\ \bibnamefont {Yang}},
  \bibinfo {author} {\bibfnamefont {D.-H.}\ \bibnamefont {Kim}}, \bibinfo
  {author} {\bibfnamefont {J.}~\bibnamefont {Byun}}, \bibinfo {author}
  {\bibfnamefont {W.-L.}\ \bibnamefont {Xu}}, \bibinfo {author} {\bibfnamefont
  {P.~R.}\ \bibnamefont {Nicovich}}, \bibinfo {author} {\bibfnamefont
  {R.}~\bibnamefont {Cheong}}, \bibinfo {author} {\bibfnamefont
  {A.}~\bibnamefont {Levchenko}}, \bibinfo {author} {\bibfnamefont
  {N.}~\bibnamefont {Thakor}}, \bibinfo {author} {\bibfnamefont {F.-Q.}\
  \bibnamefont {Zhou}},  \emph {et~al.},\ }\href@noop {} {\bibfield  {journal}
  {\bibinfo  {journal} {Proceedings of the National Academy of Sciences}\
  }\textbf {\bibinfo {volume} {108}},\ \bibinfo {pages} {5057} (\bibinfo {year}
  {2011})}\BibitemShut {NoStop}%
\bibitem [{\citenamefont {Zheng}\ \emph {et~al.}(1996)\citenamefont {Zheng},
  \citenamefont {Wan},\ and\ \citenamefont {Poo}}]{Zheng1996}%
  \BibitemOpen
  \bibfield  {author} {\bibinfo {author} {\bibfnamefont {J.~Q.}\ \bibnamefont
  {Zheng}}, \bibinfo {author} {\bibfnamefont {J.}~\bibnamefont {Wan}}, \ and\
  \bibinfo {author} {\bibfnamefont {M.}~\bibnamefont {Poo}},\ }\href@noop {}
  {\bibfield  {journal} {\bibinfo  {journal} {The Journal of neuroscience}\
  }\textbf {\bibinfo {volume} {16}},\ \bibinfo {pages} {1140} (\bibinfo {year}
  {1996})}\BibitemShut {NoStop}%
\bibitem [{\citenamefont {Dennerll}\ \emph {et~al.}(1989)\citenamefont
  {Dennerll}, \citenamefont {Lamoureux}, \citenamefont {Buxbaum},\ and\
  \citenamefont {Heidemann}}]{Dennerll1989}%
  \BibitemOpen
  \bibfield  {author} {\bibinfo {author} {\bibfnamefont {T.~J.}\ \bibnamefont
  {Dennerll}}, \bibinfo {author} {\bibfnamefont {P.}~\bibnamefont {Lamoureux}},
  \bibinfo {author} {\bibfnamefont {R.~E.}\ \bibnamefont {Buxbaum}}, \ and\
  \bibinfo {author} {\bibfnamefont {S.~R.}\ \bibnamefont {Heidemann}},\
  }\href@noop {} {\bibfield  {journal} {\bibinfo  {journal} {The journal of
  cell biology}\ }\textbf {\bibinfo {volume} {109}},\ \bibinfo {pages} {3073}
  (\bibinfo {year} {1989})}\BibitemShut {NoStop}%
\bibitem [{\citenamefont {Bernal}\ \emph {et~al.}(2007)\citenamefont {Bernal},
  \citenamefont {Pullarkat},\ and\ \citenamefont {Melo}}]{Bernal2007}%
  \BibitemOpen
  \bibfield  {author} {\bibinfo {author} {\bibfnamefont {R.}~\bibnamefont
  {Bernal}}, \bibinfo {author} {\bibfnamefont {P.~A.}\ \bibnamefont
  {Pullarkat}}, \ and\ \bibinfo {author} {\bibfnamefont {F.}~\bibnamefont
  {Melo}},\ }\href@noop {} {\bibfield  {journal} {\bibinfo  {journal} {Physical
  review letters}\ }\textbf {\bibinfo {volume} {99}},\ \bibinfo {pages}
  {018301} (\bibinfo {year} {2007})}\BibitemShut {NoStop}%
\bibitem [{\citenamefont {Bernal}\ \emph {et~al.}(2010)\citenamefont {Bernal},
  \citenamefont {Melo},\ and\ \citenamefont {Pullarkat}}]{Bernal2010}%
  \BibitemOpen
  \bibfield  {author} {\bibinfo {author} {\bibfnamefont {R.}~\bibnamefont
  {Bernal}}, \bibinfo {author} {\bibfnamefont {F.}~\bibnamefont {Melo}}, \ and\
  \bibinfo {author} {\bibfnamefont {P.~A.}\ \bibnamefont {Pullarkat}},\
  }\href@noop {} {\bibfield  {journal} {\bibinfo  {journal} {Biophysical
  journal}\ }\textbf {\bibinfo {volume} {98}},\ \bibinfo {pages} {515}
  (\bibinfo {year} {2010})}\BibitemShut {NoStop}%
\bibitem [{\citenamefont {Zheng}\ \emph {et~al.}(1991)\citenamefont {Zheng},
  \citenamefont {Lamoureux}, \citenamefont {Santiago}, \citenamefont
  {Dennerll}, \citenamefont {Buxbaum},\ and\ \citenamefont
  {Heidemann}}]{Zheng1991}%
  \BibitemOpen
  \bibfield  {author} {\bibinfo {author} {\bibfnamefont {J.}~\bibnamefont
  {Zheng}}, \bibinfo {author} {\bibfnamefont {P.}~\bibnamefont {Lamoureux}},
  \bibinfo {author} {\bibfnamefont {V.}~\bibnamefont {Santiago}}, \bibinfo
  {author} {\bibfnamefont {T.}~\bibnamefont {Dennerll}}, \bibinfo {author}
  {\bibfnamefont {R.~E.}\ \bibnamefont {Buxbaum}}, \ and\ \bibinfo {author}
  {\bibfnamefont {S.~R.}\ \bibnamefont {Heidemann}},\ }\href@noop {} {\bibfield
   {journal} {\bibinfo  {journal} {The Journal of neuroscience}\ }\textbf
  {\bibinfo {volume} {11}},\ \bibinfo {pages} {1117} (\bibinfo {year}
  {1991})}\BibitemShut {NoStop}%
\bibitem [{\citenamefont {J{\'e}rusalem}\ \emph {et~al.}(2014)\citenamefont
  {J{\'e}rusalem}, \citenamefont {Garc{\'\i}a-Grajales}, \citenamefont
  {Merch{\'a}n-P{\'e}rez},\ and\ \citenamefont {Pe{\~n}a}}]{Jerusalem2014}%
  \BibitemOpen
  \bibfield  {author} {\bibinfo {author} {\bibfnamefont {A.}~\bibnamefont
  {J{\'e}rusalem}}, \bibinfo {author} {\bibfnamefont {J.~A.}\ \bibnamefont
  {Garc{\'\i}a-Grajales}}, \bibinfo {author} {\bibfnamefont {A.}~\bibnamefont
  {Merch{\'a}n-P{\'e}rez}}, \ and\ \bibinfo {author} {\bibfnamefont {J.~M.}\
  \bibnamefont {Pe{\~n}a}},\ }\href@noop {} {\bibfield  {journal} {\bibinfo
  {journal} {Biomechanics and modeling in mechanobiology}\ }\textbf {\bibinfo
  {volume} {13}},\ \bibinfo {pages} {883} (\bibinfo {year} {2014})}\BibitemShut
  {NoStop}%
\bibitem [{\citenamefont {Ahmed}\ and\ \citenamefont {Saif}(2014)}]{Ahmed2014}%
  \BibitemOpen
  \bibfield  {author} {\bibinfo {author} {\bibfnamefont {W.~W.}\ \bibnamefont
  {Ahmed}}\ and\ \bibinfo {author} {\bibfnamefont {T.~A.}\ \bibnamefont
  {Saif}},\ }\href@noop {} {\bibfield  {journal} {\bibinfo  {journal}
  {Scientific reports}\ }\textbf {\bibinfo {volume} {4}},\ \bibinfo {pages}
  {4481} (\bibinfo {year} {2014})}\BibitemShut {NoStop}%
\bibitem [{\citenamefont {Shamloo}\ \emph {et~al.}(2015)\citenamefont
  {Shamloo}, \citenamefont {Manuchehrfar},\ and\ \citenamefont
  {Rafii-Tabar}}]{Shamloo2015}%
  \BibitemOpen
  \bibfield  {author} {\bibinfo {author} {\bibfnamefont {A.}~\bibnamefont
  {Shamloo}}, \bibinfo {author} {\bibfnamefont {F.}~\bibnamefont
  {Manuchehrfar}}, \ and\ \bibinfo {author} {\bibfnamefont {H.}~\bibnamefont
  {Rafii-Tabar}},\ }\href@noop {} {\bibfield  {journal} {\bibinfo  {journal}
  {Journal of biomechanics}\ }\textbf {\bibinfo {volume} {48}},\ \bibinfo
  {pages} {1241} (\bibinfo {year} {2015})}\BibitemShut {NoStop}%
\bibitem [{\citenamefont {Betz}\ \emph {et~al.}(2011)\citenamefont {Betz},
  \citenamefont {Koch}, \citenamefont {Lu}, \citenamefont {Franze},\ and\
  \citenamefont {K{\"a}s}}]{Betz2011}%
  \BibitemOpen
  \bibfield  {author} {\bibinfo {author} {\bibfnamefont {T.}~\bibnamefont
  {Betz}}, \bibinfo {author} {\bibfnamefont {D.}~\bibnamefont {Koch}}, \bibinfo
  {author} {\bibfnamefont {Y.-B.}\ \bibnamefont {Lu}}, \bibinfo {author}
  {\bibfnamefont {K.}~\bibnamefont {Franze}}, \ and\ \bibinfo {author}
  {\bibfnamefont {J.~A.}\ \bibnamefont {K{\"a}s}},\ }\href@noop {} {\bibfield
  {journal} {\bibinfo  {journal} {Proceedings of the National Academy of
  Sciences}\ }\textbf {\bibinfo {volume} {108}},\ \bibinfo {pages} {13420}
  (\bibinfo {year} {2011})}\BibitemShut {NoStop}%
\bibitem [{\citenamefont {Betz}\ \emph {et~al.}(2006)\citenamefont {Betz},
  \citenamefont {Lim},\ and\ \citenamefont {K{\"a}s}}]{Betz2006}%
  \BibitemOpen
  \bibfield  {author} {\bibinfo {author} {\bibfnamefont {T.}~\bibnamefont
  {Betz}}, \bibinfo {author} {\bibfnamefont {D.}~\bibnamefont {Lim}}, \ and\
  \bibinfo {author} {\bibfnamefont {J.~A.}\ \bibnamefont {K{\"a}s}},\
  }\href@noop {} {\bibfield  {journal} {\bibinfo  {journal} {Physical review
  letters}\ }\textbf {\bibinfo {volume} {96}},\ \bibinfo {pages} {098103}
  (\bibinfo {year} {2006})}\BibitemShut {NoStop}%
\bibitem [{\citenamefont {Craig}\ \emph {et~al.}(2012)\citenamefont {Craig},
  \citenamefont {Van~Goor}, \citenamefont {Forscher},\ and\ \citenamefont
  {Mogilner}}]{Craig2012}%
  \BibitemOpen
  \bibfield  {author} {\bibinfo {author} {\bibfnamefont {E.~M.}\ \bibnamefont
  {Craig}}, \bibinfo {author} {\bibfnamefont {D.}~\bibnamefont {Van~Goor}},
  \bibinfo {author} {\bibfnamefont {P.}~\bibnamefont {Forscher}}, \ and\
  \bibinfo {author} {\bibfnamefont {A.}~\bibnamefont {Mogilner}},\ }\href@noop
  {} {\bibfield  {journal} {\bibinfo  {journal} {Biophysical journal}\ }\textbf
  {\bibinfo {volume} {102}},\ \bibinfo {pages} {1503} (\bibinfo {year}
  {2012})}\BibitemShut {NoStop}%
\bibitem [{\citenamefont {Lamoureux}\ \emph {et~al.}(2002)\citenamefont
  {Lamoureux}, \citenamefont {Ruthel}, \citenamefont {Buxbaum},\ and\
  \citenamefont {Heidemann}}]{Lamoureux2002}%
  \BibitemOpen
  \bibfield  {author} {\bibinfo {author} {\bibfnamefont {P.}~\bibnamefont
  {Lamoureux}}, \bibinfo {author} {\bibfnamefont {G.}~\bibnamefont {Ruthel}},
  \bibinfo {author} {\bibfnamefont {R.~E.}\ \bibnamefont {Buxbaum}}, \ and\
  \bibinfo {author} {\bibfnamefont {S.~R.}\ \bibnamefont {Heidemann}},\
  }\href@noop {} {\bibfield  {journal} {\bibinfo  {journal} {The Journal of
  cell biology}\ }\textbf {\bibinfo {volume} {159}},\ \bibinfo {pages} {499}
  (\bibinfo {year} {2002})}\BibitemShut {NoStop}%
\bibitem [{\citenamefont {Nguyen}\ \emph {et~al.}(2013)\citenamefont {Nguyen},
  \citenamefont {Hogue}, \citenamefont {Cung}, \citenamefont {Purohit},\ and\
  \citenamefont {McAlpine}}]{Nguyen2013}%
  \BibitemOpen
  \bibfield  {author} {\bibinfo {author} {\bibfnamefont {T.~D.}\ \bibnamefont
  {Nguyen}}, \bibinfo {author} {\bibfnamefont {I.~B.}\ \bibnamefont {Hogue}},
  \bibinfo {author} {\bibfnamefont {K.}~\bibnamefont {Cung}}, \bibinfo {author}
  {\bibfnamefont {P.~K.}\ \bibnamefont {Purohit}}, \ and\ \bibinfo {author}
  {\bibfnamefont {M.~C.}\ \bibnamefont {McAlpine}},\ }\href@noop {} {\bibfield
  {journal} {\bibinfo  {journal} {Lab on a Chip}\ }\textbf {\bibinfo {volume}
  {13}},\ \bibinfo {pages} {3735} (\bibinfo {year} {2013})}\BibitemShut
  {NoStop}%
\bibitem [{\citenamefont {Heidemann}\ and\ \citenamefont
  {Buxbaum}(1993)}]{Heidemann1993}%
  \BibitemOpen
  \bibfield  {author} {\bibinfo {author} {\bibfnamefont {S.~R.}\ \bibnamefont
  {Heidemann}}\ and\ \bibinfo {author} {\bibfnamefont {R.~E.}\ \bibnamefont
  {Buxbaum}},\ }\href@noop {} {\bibfield  {journal} {\bibinfo  {journal}
  {Neurotoxicology}\ }\textbf {\bibinfo {volume} {15}},\ \bibinfo {pages} {95}
  (\bibinfo {year} {1993})}\BibitemShut {NoStop}%
\bibitem [{\citenamefont {Ahmad}\ \emph {et~al.}(2000)\citenamefont {Ahmad},
  \citenamefont {Hughey}, \citenamefont {Wittmann}, \citenamefont {Hyman},
  \citenamefont {Greaser},\ and\ \citenamefont {Baas}}]{Ahmad2000}%
  \BibitemOpen
  \bibfield  {author} {\bibinfo {author} {\bibfnamefont {F.~J.}\ \bibnamefont
  {Ahmad}}, \bibinfo {author} {\bibfnamefont {J.}~\bibnamefont {Hughey}},
  \bibinfo {author} {\bibfnamefont {T.}~\bibnamefont {Wittmann}}, \bibinfo
  {author} {\bibfnamefont {A.}~\bibnamefont {Hyman}}, \bibinfo {author}
  {\bibfnamefont {M.}~\bibnamefont {Greaser}}, \ and\ \bibinfo {author}
  {\bibfnamefont {P.~W.}\ \bibnamefont {Baas}},\ }\href@noop {} {\bibfield
  {journal} {\bibinfo  {journal} {Nature cell biology}\ }\textbf {\bibinfo
  {volume} {2}},\ \bibinfo {pages} {276} (\bibinfo {year} {2000})}\BibitemShut
  {NoStop}%
\bibitem [{\citenamefont {Ruschel}\ \emph {et~al.}(2015)\citenamefont
  {Ruschel}, \citenamefont {Hellal}, \citenamefont {Flynn}, \citenamefont
  {Dupraz}, \citenamefont {Elliott}, \citenamefont {Tedeschi}, \citenamefont
  {Bates}, \citenamefont {Sliwinski}, \citenamefont {Brook}, \citenamefont
  {Dobrindt} \emph {et~al.}}]{ruschel2015systemic}%
  \BibitemOpen
  \bibfield  {author} {\bibinfo {author} {\bibfnamefont {J.}~\bibnamefont
  {Ruschel}}, \bibinfo {author} {\bibfnamefont {F.}~\bibnamefont {Hellal}},
  \bibinfo {author} {\bibfnamefont {K.~C.}\ \bibnamefont {Flynn}}, \bibinfo
  {author} {\bibfnamefont {S.}~\bibnamefont {Dupraz}}, \bibinfo {author}
  {\bibfnamefont {D.~A.}\ \bibnamefont {Elliott}}, \bibinfo {author}
  {\bibfnamefont {A.}~\bibnamefont {Tedeschi}}, \bibinfo {author}
  {\bibfnamefont {M.}~\bibnamefont {Bates}}, \bibinfo {author} {\bibfnamefont
  {C.}~\bibnamefont {Sliwinski}}, \bibinfo {author} {\bibfnamefont
  {G.}~\bibnamefont {Brook}}, \bibinfo {author} {\bibfnamefont
  {K.}~\bibnamefont {Dobrindt}},  \emph {et~al.},\ }\href@noop {} {\bibfield
  {journal} {\bibinfo  {journal} {Science}\ }\textbf {\bibinfo {volume}
  {348}},\ \bibinfo {pages} {347} (\bibinfo {year} {2015})}\BibitemShut
  {NoStop}%
\bibitem [{\citenamefont {Yu}\ \emph {et~al.}(2012)\citenamefont {Yu},
  \citenamefont {Santiago}, \citenamefont {Katagiri},\ and\ \citenamefont
  {Geller}}]{Yu2012}%
  \BibitemOpen
  \bibfield  {author} {\bibinfo {author} {\bibfnamefont {P.}~\bibnamefont
  {Yu}}, \bibinfo {author} {\bibfnamefont {L.~Y.}\ \bibnamefont {Santiago}},
  \bibinfo {author} {\bibfnamefont {Y.}~\bibnamefont {Katagiri}}, \ and\
  \bibinfo {author} {\bibfnamefont {H.~M.}\ \bibnamefont {Geller}},\
  }\href@noop {} {\bibfield  {journal} {\bibinfo  {journal} {Journal of
  neurochemistry}\ }\textbf {\bibinfo {volume} {120}},\ \bibinfo {pages} {1117}
  (\bibinfo {year} {2012})}\BibitemShut {NoStop}%
\bibitem [{\citenamefont {Ketschek}\ \emph {et~al.}(2007)\citenamefont
  {Ketschek}, \citenamefont {Jones},\ and\ \citenamefont
  {Gallo}}]{Ketschek2007}%
  \BibitemOpen
  \bibfield  {author} {\bibinfo {author} {\bibfnamefont {A.~R.}\ \bibnamefont
  {Ketschek}}, \bibinfo {author} {\bibfnamefont {S.~L.}\ \bibnamefont {Jones}},
  \ and\ \bibinfo {author} {\bibfnamefont {G.}~\bibnamefont {Gallo}},\
  }\href@noop {} {\bibfield  {journal} {\bibinfo  {journal} {Developmental
  neurobiology}\ }\textbf {\bibinfo {volume} {67}},\ \bibinfo {pages} {1305}
  (\bibinfo {year} {2007})}\BibitemShut {NoStop}%
\bibitem [{\citenamefont {Holland}\ \emph {et~al.}(2015)\citenamefont
  {Holland}, \citenamefont {Miller},\ and\ \citenamefont {Kuhl}}]{Holland2014}%
  \BibitemOpen
  \bibfield  {author} {\bibinfo {author} {\bibfnamefont {M.~A.}\ \bibnamefont
  {Holland}}, \bibinfo {author} {\bibfnamefont {K.~E.}\ \bibnamefont {Miller}},
  \ and\ \bibinfo {author} {\bibfnamefont {E.}~\bibnamefont {Kuhl}},\
  }\href@noop {} {\bibfield  {journal} {\bibinfo  {journal} {Annals of
  biomedical engineering}\ }\textbf {\bibinfo {volume} {43}},\ \bibinfo {pages}
  {1640} (\bibinfo {year} {2015})}\BibitemShut {NoStop}%
\bibitem [{\citenamefont {Prost}\ \emph {et~al.}(2015)\citenamefont {Prost},
  \citenamefont {J{\"u}licher},\ and\ \citenamefont
  {Joanny}}]{prost2015active}%
  \BibitemOpen
  \bibfield  {author} {\bibinfo {author} {\bibfnamefont {J.}~\bibnamefont
  {Prost}}, \bibinfo {author} {\bibfnamefont {F.}~\bibnamefont {J{\"u}licher}},
  \ and\ \bibinfo {author} {\bibfnamefont {J.}~\bibnamefont {Joanny}},\
  }\href@noop {} {\bibfield  {journal} {\bibinfo  {journal} {Nature Physics}\
  }\textbf {\bibinfo {volume} {11}},\ \bibinfo {pages} {111} (\bibinfo {year}
  {2015})}\BibitemShut {NoStop}%
\bibitem [{\citenamefont {Bressloff}\ and\ \citenamefont
  {Levien}(2015)}]{Bressloff2015}%
  \BibitemOpen
  \bibfield  {author} {\bibinfo {author} {\bibfnamefont {P.~C.}\ \bibnamefont
  {Bressloff}}\ and\ \bibinfo {author} {\bibfnamefont {E.}~\bibnamefont
  {Levien}},\ }\href@noop {} {\bibfield  {journal} {\bibinfo  {journal}
  {Physical review letters}\ }\textbf {\bibinfo {volume} {114}},\ \bibinfo
  {pages} {168101} (\bibinfo {year} {2015})}\BibitemShut {NoStop}%
\bibitem [{\citenamefont {Recho}\ \emph {et~al.}(2013)\citenamefont {Recho},
  \citenamefont {Putelat},\ and\ \citenamefont {Truskinovsky}}]{Recho2013b}%
  \BibitemOpen
  \bibfield  {author} {\bibinfo {author} {\bibfnamefont {P.}~\bibnamefont
  {Recho}}, \bibinfo {author} {\bibfnamefont {T.}~\bibnamefont {Putelat}}, \
  and\ \bibinfo {author} {\bibfnamefont {L.}~\bibnamefont {Truskinovsky}},\
  }\href@noop {} {\bibfield  {journal} {\bibinfo  {journal} {Physical review
  letters}\ }\textbf {\bibinfo {volume} {111}},\ \bibinfo {pages} {108102}
  (\bibinfo {year} {2013})}\BibitemShut {NoStop}%
\bibitem [{\citenamefont {Tawada}\ and\ \citenamefont
  {Sekimoto}(1991)}]{Tawada1991}%
  \BibitemOpen
  \bibfield  {author} {\bibinfo {author} {\bibfnamefont {K.}~\bibnamefont
  {Tawada}}\ and\ \bibinfo {author} {\bibfnamefont {K.}~\bibnamefont
  {Sekimoto}},\ }\href@noop {} {\bibfield  {journal} {\bibinfo  {journal}
  {Journal of theoretical biology}\ }\textbf {\bibinfo {volume} {150}},\
  \bibinfo {pages} {193} (\bibinfo {year} {1991})}\BibitemShut {NoStop}%
\bibitem [{\citenamefont {Kruse}\ \emph {et~al.}(2005)\citenamefont {Kruse},
  \citenamefont {Joanny}, \citenamefont {J{\"u}licher}, \citenamefont {Prost},\
  and\ \citenamefont {Sekimoto}}]{Kruse2005}%
  \BibitemOpen
  \bibfield  {author} {\bibinfo {author} {\bibfnamefont {K.}~\bibnamefont
  {Kruse}}, \bibinfo {author} {\bibfnamefont {J.-F.}\ \bibnamefont {Joanny}},
  \bibinfo {author} {\bibfnamefont {F.}~\bibnamefont {J{\"u}licher}}, \bibinfo
  {author} {\bibfnamefont {J.}~\bibnamefont {Prost}}, \ and\ \bibinfo {author}
  {\bibfnamefont {K.}~\bibnamefont {Sekimoto}},\ }\href@noop {} {\bibfield
  {journal} {\bibinfo  {journal} {The European Physical Journal E}\ }\textbf
  {\bibinfo {volume} {16}},\ \bibinfo {pages} {5} (\bibinfo {year}
  {2005})}\BibitemShut {NoStop}%
\bibitem [{\citenamefont {Hannezo}\ \emph {et~al.}(2015)\citenamefont
  {Hannezo}, \citenamefont {Dong}, \citenamefont {Recho}, \citenamefont
  {Joanny},\ and\ \citenamefont {Hayashi}}]{Hannezo2015}%
  \BibitemOpen
  \bibfield  {author} {\bibinfo {author} {\bibfnamefont {E.}~\bibnamefont
  {Hannezo}}, \bibinfo {author} {\bibfnamefont {B.}~\bibnamefont {Dong}},
  \bibinfo {author} {\bibfnamefont {P.}~\bibnamefont {Recho}}, \bibinfo
  {author} {\bibfnamefont {J.-F.}\ \bibnamefont {Joanny}}, \ and\ \bibinfo
  {author} {\bibfnamefont {S.}~\bibnamefont {Hayashi}},\ }\href@noop {}
  {\bibfield  {journal} {\bibinfo  {journal} {Proceedings of the National
  Academy of Sciences}\ }\textbf {\bibinfo {volume} {112}},\ \bibinfo {pages}
  {8620} (\bibinfo {year} {2015})}\BibitemShut {NoStop}%
\bibitem [{\citenamefont {Tomba}\ \emph {et~al.}(2014)\citenamefont {Tomba},
  \citenamefont {Braini}, \citenamefont {Wu}, \citenamefont {Gov},\ and\
  \citenamefont {Villard}}]{Tomba2014}%
  \BibitemOpen
  \bibfield  {author} {\bibinfo {author} {\bibfnamefont {C.}~\bibnamefont
  {Tomba}}, \bibinfo {author} {\bibfnamefont {C.}~\bibnamefont {Braini}},
  \bibinfo {author} {\bibfnamefont {B.}~\bibnamefont {Wu}}, \bibinfo {author}
  {\bibfnamefont {N.~S.}\ \bibnamefont {Gov}}, \ and\ \bibinfo {author}
  {\bibfnamefont {C.}~\bibnamefont {Villard}},\ }\href {\doibase
  10.1039/C3SM52342J} {\bibfield  {journal} {\bibinfo  {journal} {Soft Matter}\
  }\textbf {\bibinfo {volume} {10}},\ \bibinfo {pages} {2381} (\bibinfo {year}
  {2014})}\BibitemShut {NoStop}%
\bibitem [{\citenamefont {Carlsson}(2011)}]{Carlsson2011}%
  \BibitemOpen
  \bibfield  {author} {\bibinfo {author} {\bibfnamefont {A.}~\bibnamefont
  {Carlsson}},\ }\href@noop {} {\bibfield  {journal} {\bibinfo  {journal} {New
  journal of physics}\ }\textbf {\bibinfo {volume} {13}},\ \bibinfo {pages}
  {073009} (\bibinfo {year} {2011})}\BibitemShut {NoStop}%
\bibitem [{\citenamefont {Recho}\ and\ \citenamefont
  {Truskinovsky}(2015)}]{Recho2015a}%
  \BibitemOpen
  \bibfield  {author} {\bibinfo {author} {\bibfnamefont {P.}~\bibnamefont
  {Recho}}\ and\ \bibinfo {author} {\bibfnamefont {L.}~\bibnamefont
  {Truskinovsky}},\ }\href {\doibase 10.1177/1081286515588675} {\bibfield
  {journal} {\bibinfo  {journal} {Mathematics and Mechanics of Solids}\ }
  (\bibinfo {year} {2015}),\ 10.1177/1081286515588675}\BibitemShut {NoStop}%
\bibitem [{\citenamefont {Recho}\ \emph {et~al.}(2015)\citenamefont {Recho},
  \citenamefont {Putelat},\ and\ \citenamefont {Truskinovsky}}]{Recho2015b}%
  \BibitemOpen
  \bibfield  {author} {\bibinfo {author} {\bibfnamefont {P.}~\bibnamefont
  {Recho}}, \bibinfo {author} {\bibfnamefont {T.}~\bibnamefont {Putelat}}, \
  and\ \bibinfo {author} {\bibfnamefont {L.}~\bibnamefont {Truskinovsky}},\
  }\href@noop {} {\bibfield  {journal} {\bibinfo  {journal} {Journal of the
  Mechanics and Physics of Solids}\ }\textbf {\bibinfo {volume} {84}},\
  \bibinfo {pages} {469} (\bibinfo {year} {2015})}\BibitemShut {NoStop}%
\bibitem [{\citenamefont {Xu}\ \emph {et~al.}(2013)\citenamefont {Xu},
  \citenamefont {Zhong},\ and\ \citenamefont {Zhuang}}]{Xu2013}%
  \BibitemOpen
  \bibfield  {author} {\bibinfo {author} {\bibfnamefont {K.}~\bibnamefont
  {Xu}}, \bibinfo {author} {\bibfnamefont {G.}~\bibnamefont {Zhong}}, \ and\
  \bibinfo {author} {\bibfnamefont {X.}~\bibnamefont {Zhuang}},\ }\href@noop {}
  {\bibfield  {journal} {\bibinfo  {journal} {Science}\ }\textbf {\bibinfo
  {volume} {339}},\ \bibinfo {pages} {452} (\bibinfo {year}
  {2013})}\BibitemShut {NoStop}%
\bibitem [{\citenamefont {Moulton}\ \emph {et~al.}(2012)\citenamefont
  {Moulton}, \citenamefont {Lessinnes},\ and\ \citenamefont
  {Goriely}}]{molego12}%
  \BibitemOpen
  \bibfield  {author} {\bibinfo {author} {\bibfnamefont {D.~E.}\ \bibnamefont
  {Moulton}}, \bibinfo {author} {\bibfnamefont {T.}~\bibnamefont {Lessinnes}},
  \ and\ \bibinfo {author} {\bibfnamefont {A.}~\bibnamefont {Goriely}},\
  }\href@noop {} {\bibfield  {journal} {\bibinfo  {journal} {Journal of the
  Mechanics and Physics of Solids}\ }\textbf {\bibinfo {volume} {61}},\
  \bibinfo {pages} {398} (\bibinfo {year} {2012})}\BibitemShut {NoStop}%
\bibitem [{\citenamefont {Rubinstein}\ \emph {et~al.}(2009)\citenamefont
  {Rubinstein}, \citenamefont {Fournier}, \citenamefont {Jacobson},
  \citenamefont {Verkhovsky},\ and\ \citenamefont {Mogilner}}]{Rubinstein2009}%
  \BibitemOpen
  \bibfield  {author} {\bibinfo {author} {\bibfnamefont {B.}~\bibnamefont
  {Rubinstein}}, \bibinfo {author} {\bibfnamefont {M.~F.}\ \bibnamefont
  {Fournier}}, \bibinfo {author} {\bibfnamefont {K.}~\bibnamefont {Jacobson}},
  \bibinfo {author} {\bibfnamefont {A.~B.}\ \bibnamefont {Verkhovsky}}, \ and\
  \bibinfo {author} {\bibfnamefont {A.}~\bibnamefont {Mogilner}},\ }\href@noop
  {} {\bibfield  {journal} {\bibinfo  {journal} {Biophysical journal}\ }\textbf
  {\bibinfo {volume} {97}},\ \bibinfo {pages} {1853} (\bibinfo {year}
  {2009})}\BibitemShut {NoStop}%
\bibitem [{\citenamefont {Recho}\ and\ \citenamefont
  {Truskinovsky}(2013)}]{Recho2013a}%
  \BibitemOpen
  \bibfield  {author} {\bibinfo {author} {\bibfnamefont {P.}~\bibnamefont
  {Recho}}\ and\ \bibinfo {author} {\bibfnamefont {L.}~\bibnamefont
  {Truskinovsky}},\ }\href@noop {} {\bibfield  {journal} {\bibinfo  {journal}
  {Physical Review E}\ }\textbf {\bibinfo {volume} {87}},\ \bibinfo {pages}
  {022720} (\bibinfo {year} {2013})}\BibitemShut {NoStop}%
\bibitem [{\citenamefont {Jiang}\ and\ \citenamefont {Sun}(2013)}]{Jiang2013}%
  \BibitemOpen
  \bibfield  {author} {\bibinfo {author} {\bibfnamefont {H.}~\bibnamefont
  {Jiang}}\ and\ \bibinfo {author} {\bibfnamefont {S.~X.}\ \bibnamefont
  {Sun}},\ }\href@noop {} {\bibfield  {journal} {\bibinfo  {journal}
  {Biophysical journal}\ }\textbf {\bibinfo {volume} {105}},\ \bibinfo {pages}
  {609} (\bibinfo {year} {2013})}\BibitemShut {NoStop}%
\bibitem [{\citenamefont {Hui}\ \emph {et~al.}(2014)\citenamefont {Hui},
  \citenamefont {Zhou}, \citenamefont {Qian}, \citenamefont {Lin},
  \citenamefont {Ngan},\ and\ \citenamefont {Gao}}]{Hui2014}%
  \BibitemOpen
  \bibfield  {author} {\bibinfo {author} {\bibfnamefont {T.}~\bibnamefont
  {Hui}}, \bibinfo {author} {\bibfnamefont {Z.}~\bibnamefont {Zhou}}, \bibinfo
  {author} {\bibfnamefont {J.}~\bibnamefont {Qian}}, \bibinfo {author}
  {\bibfnamefont {Y.}~\bibnamefont {Lin}}, \bibinfo {author} {\bibfnamefont
  {A.}~\bibnamefont {Ngan}}, \ and\ \bibinfo {author} {\bibfnamefont
  {H.}~\bibnamefont {Gao}},\ }\href@noop {} {\bibfield  {journal} {\bibinfo
  {journal} {Physical review letters}\ }\textbf {\bibinfo {volume} {113}},\
  \bibinfo {pages} {118101} (\bibinfo {year} {2014})}\BibitemShut {NoStop}%
\bibitem [{\citenamefont {Recho}\ \emph {et~al.}(2014)\citenamefont {Recho},
  \citenamefont {Joanny},\ and\ \citenamefont {Truskinovsky}}]{Recho2014}%
  \BibitemOpen
  \bibfield  {author} {\bibinfo {author} {\bibfnamefont {P.}~\bibnamefont
  {Recho}}, \bibinfo {author} {\bibfnamefont {J.-F.}\ \bibnamefont {Joanny}}, \
  and\ \bibinfo {author} {\bibfnamefont {L.}~\bibnamefont {Truskinovsky}},\
  }\href@noop {} {\bibfield  {journal} {\bibinfo  {journal} {Physical Review
  Letters}\ }\textbf {\bibinfo {volume} {112}},\ \bibinfo {pages} {218101}
  (\bibinfo {year} {2014})}\BibitemShut {NoStop}%
\bibitem [{\citenamefont {Sayyad}\ \emph {et~al.}(2015)\citenamefont {Sayyad},
  \citenamefont {Amin}, \citenamefont {Fabris}, \citenamefont {Ercolini},\ and\
  \citenamefont {Torre}}]{Sayyad2015}%
  \BibitemOpen
  \bibfield  {author} {\bibinfo {author} {\bibfnamefont {W.~A.}\ \bibnamefont
  {Sayyad}}, \bibinfo {author} {\bibfnamefont {L.}~\bibnamefont {Amin}},
  \bibinfo {author} {\bibfnamefont {P.}~\bibnamefont {Fabris}}, \bibinfo
  {author} {\bibfnamefont {E.}~\bibnamefont {Ercolini}}, \ and\ \bibinfo
  {author} {\bibfnamefont {V.}~\bibnamefont {Torre}},\ }\href@noop {}
  {\bibfield  {journal} {\bibinfo  {journal} {Scientific reports}\ }\textbf
  {\bibinfo {volume} {5}},\ \bibinfo {pages} {7842} (\bibinfo {year}
  {2015})}\BibitemShut {NoStop}%
\bibitem [{\citenamefont {Letourneau}\ and\ \citenamefont
  {Ressler}(1984)}]{Letourneau1984}%
  \BibitemOpen
  \bibfield  {author} {\bibinfo {author} {\bibfnamefont {P.~C.}\ \bibnamefont
  {Letourneau}}\ and\ \bibinfo {author} {\bibfnamefont {A.~H.}\ \bibnamefont
  {Ressler}},\ }\href@noop {} {\bibfield  {journal} {\bibinfo  {journal} {The
  Journal of cell biology}\ }\textbf {\bibinfo {volume} {98}},\ \bibinfo
  {pages} {1355} (\bibinfo {year} {1984})}\BibitemShut {NoStop}%
\bibitem [{\citenamefont {Bamburg}\ \emph {et~al.}(1986)\citenamefont
  {Bamburg}, \citenamefont {Bray},\ and\ \citenamefont
  {Chapman}}]{Bamburg1986}%
  \BibitemOpen
  \bibfield  {author} {\bibinfo {author} {\bibfnamefont {J.}~\bibnamefont
  {Bamburg}}, \bibinfo {author} {\bibfnamefont {D.}~\bibnamefont {Bray}}, \
  and\ \bibinfo {author} {\bibfnamefont {K.}~\bibnamefont {Chapman}},\
  }\href@noop {} {\bibfield  {journal} {\bibinfo  {journal} {Nature}\ }\textbf
  {\bibinfo {volume} {321}},\ \bibinfo {pages} {788} (\bibinfo {year}
  {1986})}\BibitemShut {NoStop}%
\bibitem [{\citenamefont {Witte}\ \emph {et~al.}(2008)\citenamefont {Witte},
  \citenamefont {Neukirchen},\ and\ \citenamefont
  {Bradke}}]{witte2008microtubule}%
  \BibitemOpen
  \bibfield  {author} {\bibinfo {author} {\bibfnamefont {H.}~\bibnamefont
  {Witte}}, \bibinfo {author} {\bibfnamefont {D.}~\bibnamefont {Neukirchen}}, \
  and\ \bibinfo {author} {\bibfnamefont {F.}~\bibnamefont {Bradke}},\
  }\href@noop {} {\bibfield  {journal} {\bibinfo  {journal} {The Journal of
  cell biology}\ }\textbf {\bibinfo {volume} {180}},\ \bibinfo {pages} {619}
  (\bibinfo {year} {2008})}\BibitemShut {NoStop}%
\bibitem [{\citenamefont {Hellal}\ \emph {et~al.}(2011)\citenamefont {Hellal},
  \citenamefont {Hurtado}, \citenamefont {Ruschel}, \citenamefont {Flynn},
  \citenamefont {Laskowski}, \citenamefont {Umlauf}, \citenamefont {Kapitein},
  \citenamefont {Strikis}, \citenamefont {Lemmon}, \citenamefont {Bixby} \emph
  {et~al.}}]{hellal2011microtubule}%
  \BibitemOpen
  \bibfield  {author} {\bibinfo {author} {\bibfnamefont {F.}~\bibnamefont
  {Hellal}}, \bibinfo {author} {\bibfnamefont {A.}~\bibnamefont {Hurtado}},
  \bibinfo {author} {\bibfnamefont {J.}~\bibnamefont {Ruschel}}, \bibinfo
  {author} {\bibfnamefont {K.~C.}\ \bibnamefont {Flynn}}, \bibinfo {author}
  {\bibfnamefont {C.~J.}\ \bibnamefont {Laskowski}}, \bibinfo {author}
  {\bibfnamefont {M.}~\bibnamefont {Umlauf}}, \bibinfo {author} {\bibfnamefont
  {L.~C.}\ \bibnamefont {Kapitein}}, \bibinfo {author} {\bibfnamefont
  {D.}~\bibnamefont {Strikis}}, \bibinfo {author} {\bibfnamefont
  {V.}~\bibnamefont {Lemmon}}, \bibinfo {author} {\bibfnamefont
  {J.}~\bibnamefont {Bixby}},  \emph {et~al.},\ }\href@noop {} {\bibfield
  {journal} {\bibinfo  {journal} {Science}\ }\textbf {\bibinfo {volume}
  {331}},\ \bibinfo {pages} {928} (\bibinfo {year} {2011})}\BibitemShut
  {NoStop}%
\bibitem [{\citenamefont {Pullarkat}\ \emph {et~al.}(2006)\citenamefont
  {Pullarkat}, \citenamefont {Dommersnes}, \citenamefont {Fern{\'a}ndez},
  \citenamefont {Joanny},\ and\ \citenamefont {Ott}}]{Pullarkat2006}%
  \BibitemOpen
  \bibfield  {author} {\bibinfo {author} {\bibfnamefont {P.~A.}\ \bibnamefont
  {Pullarkat}}, \bibinfo {author} {\bibfnamefont {P.}~\bibnamefont
  {Dommersnes}}, \bibinfo {author} {\bibfnamefont {P.}~\bibnamefont
  {Fern{\'a}ndez}}, \bibinfo {author} {\bibfnamefont {J.-F.}\ \bibnamefont
  {Joanny}}, \ and\ \bibinfo {author} {\bibfnamefont {A.}~\bibnamefont {Ott}},\
  }\href@noop {} {\bibfield  {journal} {\bibinfo  {journal} {Physical review
  letters}\ }\textbf {\bibinfo {volume} {96}},\ \bibinfo {pages} {048104}
  (\bibinfo {year} {2006})}\BibitemShut {NoStop}%
\bibitem [{\citenamefont {LeVeque}(2002)}]{Leveque2002}%
  \BibitemOpen
  \bibfield  {author} {\bibinfo {author} {\bibfnamefont {R.}~\bibnamefont
  {LeVeque}},\ }\href@noop {} {\emph {\bibinfo {title} {Finite volume methods
  for hyperbolic problems}}}\ (\bibinfo  {publisher} {Cambridge University
  Press},\ \bibinfo {address} {Cambridge},\ \bibinfo {year} {2002})\BibitemShut
  {NoStop}%
\end{thebibliography}

%

\end{document}